# Discovery of New Zintl Films and Nanowires Grown by Topotaxy Conversion of III–V Semiconductors

Man Suk Song,[1,2] Lothar Houben,[3] Jean Souza,[1] Edanel Fishbein[1], Moshe Haim[1], Ambikesh Gupta,[1] Yufei Zhao,[1,4] Anna-Eden Kossoy,[3] Binghai Yan, [1,4] Haim Beidenkopf,[1]* and Hadas Shtrikman[1]*

[1]Department of Condensed Matter Physics, Weizmann Institute of Science, Rehovot 7610001, Israel

[2] Department of Semiconductor, Dong-A University, Busan 49315, Republic of Korea

[3]Department of Chemical Research Support, Weizmann Institute of Science, Rehovot, 7610001, Israel

[4] Department of Physics, The Pennsylvania State University, University Park, PA, USA

*E-mail: haim.beidenkopf@weizmann.ac.il, hadas.shtrikman@weizmann.ac.il

## ABSTRACT

**Zintl phases attract vast scientific attention thanks to their diverse structural, magnetic, thermoelectric, topological and optical properties. Recently, Zintl $Eu_3In_2As_4$ and $Eu_5In_2As_6$ nanowires with axion magneto-topology have been synthesized by molecular beam epitaxy via topotactic conversion of InAs wurtzite and zincblende nanowires, respectively. Here, we substantially extend this methodology by demonstrating that topotaxial mutual-exchange**

**growth is applicable not only to a broader set of III–V semiconductors beyond InAs but also to three-dimensional substrates whose surfaces are converted into Zintl thin films, as well as to nanowires. We report the growth of two new compounds that have not been synthesized before: $Eu_5Ga_2As_6$ thin films are converted from GaAs substrates, and $Eu_5Al_2As_6$ thin films from AlAs films. We also convert GaAs nanowires of both wurtzite and zincblende structures into $Eu_5Ga_2As_6$ nanowires. Though the stoichiometry is the same as in the previously reported $Eu_5In_2As_6$ case, the crystallographic structure of the new compounds is distinct, as microscopy and diffraction reveal a single-phase *Pnma* symmetry group rather than *Pbam*, highlighting symmetry-guided topotactic pathways to new Zintl frameworks further diversifying the ensuing phenomenology. The compounds host an intricate magnetic phase diagram with three magnetic transitions, including two distinct antiferromagnetic orders, a canted antiferromagnetic transition that evolves into another antiferromagnetic phase under applied magnetic field through a spin-flop transition. Ab initio calculations predict that both Zintls are semiconductors with gaps of 0.79 eV in $Eu_5Ga_2As_6$ and 0.90 eV in $Eu_5Al_2As_6$. The lower symmetry and increased structural *Pnma* complexity further suggest suppressed lattice thermal conductivity, pointing to thermoelectric potential alongside prospects in spintronics and detector technologies. These results outline a feasible epitaxy-compatible strategy for discovering and integrating magnetic Zintl thin films and nanowires directly from relevant III–V semiconductors.**

Discovering previously unreported compounds remains one of the enduring rewards in material science. Although synthesis and growth are often driven by specific applications, many breakthroughs have arisen from pure curiosity. Zintl phases are a prominent example of this[1,2]. They are compounds with mixed covalent and ionic bonding character, consisting of covalently bonded polyanionic frameworks – chains, nets, or rings – that are charge-balanced and stabilized by electropositive cations. The Zintl–Klemm concept originated in binary systems and has since been extended to ternary, quaternary, and even higher-order chemistries[3]. This compositional flexibility suggests the existence of numerous phases that have been theoretically predicted but not yet synthesized. The challenge in realizing these materials may stem not only from their possibly intrinsic stability but also from the limitations of conventional synthetic and crystal growth methodologies.

We have recently reported on a novel method to convert wurtzite (WZ) and zincblende (ZB) InAs nanowires (NWs) into Zintl $Eu_3In_2As_4$ and $Eu_5In_2As_6$, respectively, using topotactic mutual-exchange in molecular beam epitaxy (MBE)[4,5]. The deposition of Eu and As onto InAs (with either WZ or ZB structures) induces mutual exchanges between In from the NW core and Eu from the deposited shell. Topotactic formation of the Zintl phase is promoted by the structural similarity of the As sublattice between the initial and final compounds. The stoichiometry of the resulting ternary compound is determined by the crystal structure of the parent binary InAs; WZ InAs yields $Eu_3In_2As_4$[4], while ZB InAs yields $Eu_5In_2As_6$[5].

Building on this, this growth mechanism has the potential to be further extended both beyond NWs and to other III–V compound semiconductors and a variety of rare earth or alkaline earth metals. Here, we continue to explore it and demonstrate both the first topotactic conversion of surfaces of III-V semiconductors into Zintl matter as well as the synthesis of two novel Zintl compounds, orthorhombic $Eu_5Ga_2As_6$ and $Eu_5Al_2As_6$, which, to the best of our knowledge, have not been previously synthesized in any other growth method. We further confirmed that $Eu_5Ga_2As_6$ can be converted into both thin-film and NW forms, irrespective of the parent polytype (WZ or ZB).

As a magnetic topological material, $Eu_5In_2As_6$ and $Eu_5In_2Sb_6$, belonging to family of Zintl $Eu_{2n+1}In_2(As, Sb)_{2n+2}$, are candidates for the hybrid-order axion insulators thanks to a combined two-fold rotation with time-reversal, $C_2T$ symmetry[6–11]. Those with 5-2-6 stoichiometry crystallized in the orthorhombic *Pbam* space group. By contrast, $Eu_5Ga_2As_6$ and $Eu_5Al_2As_6$ adopt a *Pnma* space group, which is characterized by lower symmetry and greater structural complexity. In Zintl thermoelectrics (TEs), such structural complexity is associated with suppressed lattice thermal conductivity ($\kappa_l$) via enhanced phonon scattering, which can benefit the thermoelectric figure of merit $zT$[12,13]. These trends highlight space-group engineering as a useful design lever of Zintl TEs. Magnetic characterization of $Eu_5Ga_2As_6$ and $Eu_5Al_2As_6$ reveals two distinct magnetic transitions: a canted antiferromagnetic (CAFM) state at a higher temperature (~15 K) and an antiferromagnetic (AFM) state at a lower temperature (~6 K). Additionally, a spin-flop transition is observed at moderate applied magnetic fields, highlighting the complex magnetism of Zintl compounds due to their intricate crystal structure. Our work further extends the understanding of the complex magnetic competition in Zintl materials, opening an avenue for engineering complex magnetism across this broader family.

## Results and Discussions

We followed the protocol described in our previous works[4,5] and deposited Eu and As onto a ZB GaAs (001) and (111)B substrate at 520 °C (see Methods). Scanning electron microscopy (SEM) images of the resulting thin films on GaAs (001) are shown in Fig.1a (top view) and Fig.1c (side view), revealing continuous films with uniform thickness. SEM images of the thin films on GaAs (111)B are shown in Fig.S01e and Fig.S01f in Supplementary Information. To grow $Eu_5Al_2As_6$ thin films we had first to grow an AlAs buffer layer on GaAs (001), as shown in Fig.1b and Fig.1d. As demonstrated in early MBE studies[14], heteroepitaxial growth of AlAs on GaAs has been considered equivalent to homoepitaxy because of similarity in crystal structure and lattice constant. In top view SEM, both Zintl-phase films display distinct surface textures, signifying a polycrystalline structure with lateral crystallites consistent with our previous work[4,5]. The side-view SEM images clearly show the formation of Eu-Ga-As (50–110 nm thick) and Eu-Al-As (about 90 nm thick) crystalline thin films, with an AlAs buffer layer of ~15–65 nm thick in the latter.

We performed X-ray diffraction (XRD) measurements for further verification. Figure 1e and f exhibit the θ-2θ scan profiles for $Eu_5Ga_2As_6$ and $Eu_5Al_2As_6$ thin films, respectively. As direct reference data were unavailable due to the novelty of these compounds, we indexed the observed peaks using the orthorhombic $Sr_5Al_2Sb_6$-type crystal structure (space group *Pnma*) from ICSD (No. 62304) as the starting model. The lattice parameters were replaced with density functional theory (DFT)-relaxed values (Table 1), while the space group and fractional atomic coordinates were kept unchanged. Distinct diffraction peaks corresponding to each Zintl phase were clearly resolved from the GaAs substrate (and AlAs buffer layer). The $Eu_5Ga_2As_6$ film predominantly grows along its c-axis, accompanied by minor peaks assigned to the (103), (104), and (207) planes.

Likewise, the $Eu_5Al_2As_6$ films exhibit the c-axis preferred orientation. These XRD results provide macroscopic confirmation of the crystal structure across the film.

The crystal structure was, in fact, first identified at the atomic scale. We prepared cross-sectional lamellae using the focused ion beam (FIB) technique and performed high-resolution high-angle annular dark-field scanning transmission electron microscopy (HAADF-STEM) imaging (Fig. 2a–d), which revealed the $Sr_5Al_2Sb_6$-type structure directly. Our previous work established that the stoichiometry of the resulting topotactic phase is governed by the crystal structure of the host: growing Eu and As on WZ-InAs yielded $Eu_3In_2As_4$ (space group *Pnnm*), whereas growth on ZB-InAs resulted in $Eu_5In_2As_6$ (space group *Pbam*). They both belong to the Zintl family $Eu_{2n+1}M_2Pn_{2n}$, which forms narrow-gap semiconductors or semimetals. Based on this precedent, we anticipated that the topotactic reaction with ZB-GaAs would yield the isostructural $Eu_5Ga_2As_6$. This expectation was partly confirmed by energy-dispersive X-ray spectroscopy (EDS) analysis (Figure S02 h). However, despite the identical stoichiometry, we found that the material crystallized in the *Pnma* space group, adopting the $Sr_5Al_2Sb_6$-type structure. They thus belong to a distinct Zintl family $Eu_{2n+1}M_nPn_{3n}$ that electronically form magnetic large-gap semiconductors. Notably, neither $Eu_5Ga_2As_6$ nor its aluminum analogue $Eu_5Al_2As_6$ has been reported in the literature. Thus, to the best of our knowledge, this work presents the first experimental realization and structural identification of these novel ternary compounds.

Atomic-resolution STEM data for the new Zintl phases $Eu_5Ga_2As_6$ and $Eu_5Al_2As_6$, grown by the topotaxial transformation of GaAs and AlAs wafers, are displayed in Fig. 2. Cross-sectional views of the transformation interface between $Eu_5Ga_2As_6$ and the GaAs (111)B substrate are shown in Fig. 2a and $Eu_5Al_2As_6$ on AlAs (100) buffer layer in Fig. 2b. A sharply defined two-dimensional transformation boundary extending over only a few nanometers in depth is observed on these substrate orientations for both Ga and As. On top of the nanometer-sharp transformation boundary, single-crystal Zintl domains up to a few 100 nm wide extend to the full thickness of the layer. We infer the atomic coordination and space group of the Zintl-phases from the atomically resolved elemental maps superimposed on the HAADF image of the atomic columns (Fig. 2c and d). Characteristic of the Zintl phases are the anionic polyhedral chains of $GaAs_4$ or $AlAs_4$ tetrahedra in charge balance with the adjacent cationic Eu. The atomic column positions and their composition along the viewing direction, either pure or mixed as in the case of Eu/As columns, are consistent with the isostructural $Sr_5Al_2Sb_6$ phase, which crystallizes in the orthorhombic space group *Pnma*[15]. The corresponding crystal structure of $Eu_5Ga_2As_6$ of $Eu_5Al_2As_6$ in the *Pnma* space group is shown in Fig. 2e and f, observed along the [100] and [010] directions of the unit cell, respectively.

The infinite anionic chains of GaAs (AlAs) are aligned in the [100] direction on the (010) plane (Fig. 2e). Each chain is formed from alternately corner- and edge-sharing $GaAs_4$ ($AlAs_4$) tetrahedra. The view in Fig. 2f corresponds with the viewing direction in the experiment in Fig. 2c and d. In line with our previous report on $Eu_5In_2As_6$[5], the topotaxial transformation progresses through the template of the polyhedral lattice nodes in the parent lattice[5]. Eu diffuses into GaAs (AlAs), occupying octahedral symmetry sites within the parent III-V lattice. The co-diffusion of Ga (Al) leads to the formation of polyhedral chains along nodes of tetrahedral lattice sites in the parent lattice that closely match.

The corner-sharing tetrahedra exist in the GaAs (AlAs) coordination in the parent phase, while edge-sharing results from Ga (Al) occupying the inversion-symmetric tetrahedral sites that are vacant in the zinc-blende parent phase. Thus, the polyanionic chains in the (010) plane of the Zintl phase align with the {110} planes in the parent lattice, where the same sequence of alternating corner and edge-shared tetrahedral sites is found. Consequently, the b axis of the Zintl phase is parallel to the three symmetry-equivalent <110> directions in the parent phase. Each of these three options allows for multiple rotations of the chains in the {110}-plane of the cubic substrate, resulting in the chains that may run in the substrate or be inclined relative to the substrate plane: within each {110} planes of the zinc-blende substrate, corner and edge-shared tetrahedral sites alternate in the same manner as in the polyanionic chains along the <112> and <110> directions. Illustration of the case where the polyanionic chains run parallel to the substrate plane is given in Fig.2, while Fig. S03 presents two possible rotations of the polyanionic chains in the (110) plane side by side. Compared with the isocompositional $Eu_5In_2As_6$ that grows in the orthorhombic *Pbam* space group when InAs is exposed to Eu [5], the relatively large chain repeat distance of four tetrahedra in the *Pnma* space group leads to a larger unit cell in the present case.

We next attempted to convert topotactically GaAs NWs as well. They have a wurtzite rather than a zincblende structure as the parental material, analogous to our approach for InAs[4]. Reclined GaAs NWs were first grown in MBE by the Au-assisted vapor–liquid–solid method on (100) and (110) GaAs substrates (see methods in SI). The core GaAs NWs were about 50 nm in diameter and 0.8–1.5 μm in length. As with the thin film growth, in the following step of exchange growth of NWs, the Ga source was closed and cooled down, and only Eu and As were supplied for 2.5 hours. SEM images of the resulting NWs grown on a (100) substrate are displayed in Fig. 3a,b and Fig. S01a,b. Two distinct morphologies were observed. As shown in Fig. 3a, most of the NWs are "short-segment" types, which bend inward toward the substrate. In contrast, Fig. 3b shows a minority of "long-segment" NWs that bend outward from the substrate and exhibit relatively larger lengths and widths compared to the former. In our previous studies, both $Eu_3In_2As_4$ and $Eu_5In_2As_6$ NWs exhibited noticeable surface roughness and a polycrystalline-like texture along the growth axis of the core NWs. Here, however, the NW surface appears comparatively smoother, and the axial grain size is larger than that of both Zintl-phase NWs transformed from InAs (Fig. 3a,b, and Fig. S01a–b).

To determine the composition of these NWs, we obtained TEM images, shown in Fig. 3. EDS spectra were acquired from dozens of regions along the NWs (Fig. S05 in SI). Unexpectedly, the phase in all was identified as $Eu_5Ga_2As_6$, both for ZB and for WZ parent structures, rather than $Eu_3Ga_2As_4$ stoichiometry as we found previously in the case of WZ InAs NWs[4]. Moreover, most NWs underwent a full exchange from the WZ NW GaAs core, regardless of their morphology. Low- and high-resolution (HR) TEM images of representative long-segment type $Eu_5Ga_2As_6$ NWs are presented in Fig. 3c and d, respectively. The middle segment in Fig. 3c, which exhibits dark diffraction contrast, shows a 34° tilt relative to the adjacent upper segment about the b-axis (Fig. S07 in SI). The HRTEM image in Figure 3d corresponds to the area marked by a red square in Fig. 3c. The Fourier transform pattern (inset), obtained from the area indicated by a white square in Fig. 3d, confirms that this segment is single-crystalline $Eu_5Ga_2As_6$ orientated along the $[\underline{3}04]$ zone axis. Notably, only a thin surface oxidation layer (less than 10 nm) is observed, with no trace of the original GaAs core, indicating complete transformation into the Zintl phase.

Meanwhile, the short-segment $Eu_5Ga_2As_6$ NWs, presented in Fig. S09, exhibit a similar structural trend: the GaAs core is almost entirely consumed except for a stem region near the substrate. Representative SEM and TEM images are displayed in Fig. S10 of the SI, together with corresponding schematics, for both long- and short-segment types. Although these two types appear morphologically distinct, both ultimately consist of fully exchanged Zintl crystallites that are joined along the growth direction. The characteristic bending observed in the short-segment NWs likely arises from the reduced length of individual segments with differing crystallographic orientations. We attribute this segmentation to the presence of stacking faults within the initial GaAs core, suggesting that the resulting grain size is dictated by the spacing between the faults. Consequently, realizing an ideal single-crystalline $Eu_5Ga_2As_6$ NW will require a growth design that suppresses stacking-fault-mediated segmentation.

As noted above, in most NWs the cores are almost completely topotactically transformed, making it difficult to capture the conversion front and clearly identify the WZ GaAs/Zintl $Eu_5Ga_2As_6$ interface. Nevertheless, Fig. 3e provides a rare TEM view of the interface between the remaining core region and the Zintl phase immediately prior to the completion of the transformation. The corresponding Fourier transform patterns from each region (white squares in Fig. 3e) are shown in Fig. 3f and g, respectively. These patterns reveal that the $(11\bar{2}0)$ plane of the WZ GaAs crystal structure is parallel to the $(010)$ plane of the Zintl phase. The nonpolar prismatic planes of WZ GaAs NWs, such as $\{\bar{1}120\}$ and $\{\bar{1}100\}$, are crystallographically analogous to the $\{110\}$ planes in a ZB GaAs substrate under the conventional ZB–WZ orientation relationship. Therefore, this topotaxial alignment of the Zintl polyanionic chains parallel to the WZ $(11\bar{2}0)$ plane is consistent with the observations in the Zintl thin film case (Fig. 2). An atomic model illustrating this interface along the viewing direction, WZ $[1\bar{1}00]$ and Zintl $[305]$ axes, is presented in Fig. 3i. These results suggest that during the topotactic formation of $Eu_5Ga_2As_6$, the polyanionic chains extend parallel to the nonpolar prismatic planes of the parent WZ GaAs NW.

The XRD θ-2θ scans of as-grown $Eu_5Ga_2As_6$ NWs (blue) and $Eu_5Ga_2As_6$ thin films (red) on GaAs (100) substrates are shown in Fig. 3h. For comparison, the thin-film spectrum–identical to that in Fig. 1e is overlaid. It contains only the $(00l)$ and $(h0l)$ reflections of $Eu_5Ga_2As_6$, together with peaks from the GaAs substrate. In principle, accurate determination of the lattice parameters of a newly identified material requires diffraction data from a randomly oriented (powder) specimen. Instead, the unique morphology and orientation of NWs can provide the necessary $(hkl)$ reflections. We measured the as-grown $Eu_5Ga_2As_6$ NWs to detect reflections with $k \neq 0$, which are absent in the textured thin film scan. The extra peaks from the NW sample, indexed as (221), (311), and (213), were utilized to determine the b-axis lattice parameter detailed in Table 1. While a discrepancy between the experimental values and DFT results is expected, given the ground-state (0 K) nature of the calculations, the overall consistency between the XRD results and the HAADF-STEM observations supports the validity of the determined lattice constants.

Next, we characterized the magnetic and electronic properties of the new compounds. For this, we have measured the magnetic H-T phase diagram of both $Eu_5Ga_2As_6$ and $Eu_5Al_2As_6$ in a commercial magnetic properties measurement system (MPMS). In both, we find rich magnetic orders with qualitatively identical sequences of transitions. They vary slightly in their exact temperature and field values. Individual magnetization traces, with the magnetic field applied parallel to the plane, measured in $Eu_5Ga_2As_6$ along temperature sweeps at varying magnetic fields, are shown in Fig. 4a. We also mapped through dM/dT in false color in Fig. 4b (the inset shows the extracted transition lines). At high temperatures and magnetic fields, we find a paramagnetic (PM) state that fits a Curie-Weiss temperature dependence (Fig. 4a, inset). The fit yields an effective moment of $\mu = 7.8\ \mu_B/Eu^{2+}$, in agreement with the divalent $4f$ $Eu^{2+}$ moment of $S = 7/2$ and L = 0, and a temperature scale of $\theta_{CW} = 2.4$ K, signifying ferromagnetic (FM) fluctuations. Indeed, a step in magnetization ΔM onsets at a temperature of $T_C$=14.7 K and vanishingly low fields. Naively, due to the small step size, it would signify a first-order transition with a weak FM order below it; however, the step feature evolves continuously into a cusp at low magnetic fields (H ~ 0.1 T). The step to cusp evolution is one of the hallmark signatures of an intrinsic easy-axis canted AFM order ($CAFM_{||EA}$)[16].

The low Curie-Weiss temperature, $\theta_{CW} << T_C$, suggests tight competition between AFM and FM fluctuations. Intriguingly, at a rather low magnetic field of about 0.7 T, there is a clear increase in the magnetization below the first transition temperature. While $\Delta M=0$ alone does not necessarily signify a second-order transition as long as a jump in entropy remains $\Delta S \neq 0$, the Clausius-Clapeyron relation yields $dT_c/dH=0$, which seems to be obeyed. Microscopically, this probably marks the transition from an intrinsic easy-axis $CAFM_{||EA}$ at low fields to a partially polarized CAFM by the external field, which polarizes the PM to a similar degree. This results in the near-vanishing of the magnetization difference between the two distinct magnetic phases. This continues until a fairly temperature-independent transition line bisects it at a yet higher field of about 0.7 T, forming a tricritical point (marked with a yellow asterisk). Above it, the low-field vertical transition line shifts progressively towards lower temperatures, signifying a second-order Neel transition ($T_{N1}$) and change into an AFM state. The rather horizontal transition comes with a positive step in magnetization, $\Delta M$, better resolved in field sweeps that are shown in Fig. 4c, thus signifying a first-order spin-flop transition where the weak FM component of the magnetization rotates from the magnetic easy-axis ($CAFM_{||EA}$) at lower fields towards the direction of the applied field ($CAFM_{||H}$). This is another signature of a CAFM state. Interestingly, low-field metamagnetic transitions have also been observed in $Eu_5In_2Sb_6$[11].

At lower temperatures, we find an additional cusp in the temperature-dependent magnetization onsetting at $T_{N2}$=5.8 K at low fields. It signifies a Neel transition to a distinct AFM-ordered state ($AFM_2$). It also intersects the spin-flop transition, $H_{SF}$, forming a second tricritical point (marked with an orange asterisk). The phase diagram of $Eu_5Al_2As_6$, plotted in dM/dT in Fig.4d (see individual curves at Fig.S16), appears essentially the same, only with slightly higher temperature and field scales. It also hosts a high-temperature canted AFM transition ($T_{N1}$=16.5 K), a low-temperature transition to a second antiferromagnetic state ($T_{N2}$=6 K), and a spin flop transition ($H_{SF}$=0.8 T) with two tricritical points. The complex magnetic structure with consecutive magnetic transitions, involving intricate magnetic structures, is characteristic of Eu-based Zintl materials such as $Eu_5In_2Sb_6$[9] and $Eu_5In_2As_6$[5].

We have further mapped the magnetic phase diagram of the $Eu_5Ga_2As_6$ NWs, as demonstrated by individual temperature sweeps in Fig. 4e (full phase diagram in Fig.S16), by dispersing them on a Si/$SiO_2$ substrate. While at low fields we identify both the 14.7 K weak FM-like cusp and the 6 K AFM transition, as in $Eu_5Ga_2As_6$ thin films, at higher fields the weak FM transition vanishes without the appearance of a second-order cusp in its absence. We also do not find a trace of the spin-flop first-order transition. In its absence, the low-temperature Neel transition, $T_{N2}$, gradually shifts to lower temperatures as the field increases, without hitting a tricritical point. These distinctions seem to arise from the lower NW dimensionality, which prevents some of the long-range magnetic correlations from developing.

After mapping accurately the stoichiometry and structure of the materials using HRTEM, EDX, and XRD, as well as their magnetic phase diagram in MPMS, we turn to predicting their band structure using density functional theory (DFT). We find remarkable agreement in the lattice constants along all three crystallographic axes among STEM, XRD, and relaxed DFT structure, summarized in Table 1. The calculated band structure of both $Eu_5Ga_2As_6$ (Fig. 4f) and $Eu_5Al_2As_6$ (Fig. 4g) is of magnetic insulators with a direct gap at the Γ point of 815 meV and 903 meV (slightly indirect gap along Γ-Z of 786 meV and 910 meV), respectively. The magnetic orders induce minor changes in the electronic dispersion. This demonstrates the ability to control the gap size of the Zintl compound by modifying the initial III-V material. This ability, within the near-infrared and telecom wavelength ranges, can serve technological applications that currently rely on doping, which inevitably introduces disorder.

A key question raised in this study is why the 5–2–6 Zintl systems—converted here both from WZ and ZB GaAs (as well as from ZB AlAs)—do not follow the parental-lattice selection rule observed in the InAs-derived system, namely WZ ⟶ 3–2–4[4] and ZB ⟶ 5–2–6[5]. Furthermore, even within the same 5–2–6 stoichiometry, the $Eu_5Ga_2As_6$ and $Eu_5Al_2As_6$ crystals adopt a different space group from $Eu_5In_2As_6$. Our observations suggest that, in the GaAs/AlAs-derived case, structural selection is not governed solely by rigid templating from the parent anion sublattice, but is strongly influenced also by steric constraints associated with A–M cation size mismatch. We therefore hypothesize a size-imbalance effect: compared to In, Ga and Al have substantially smaller covalent radii, which increases the size mismatch with the larger A-site cation (Eu) and shifts the packing preference of the ternary Zintl framework. Related cation-size effects have been discussed in ternary Zintl phases[17].

To examine whether such a steric factor can account for the observed space group selection, we compiled reported 5–2–6 Zintl phases of the form $A_5M_2Pn_6$ (A = magnetic cation; M = group 13, triel; Pn = pnictogen) and summarized their space groups in Table 2. We classify these compounds by a simple size ratio, $\# = r_A/r_M$ where $r_A$ is the ionic radius of A and $r_M$ is the covalent radius of M. In these Zintl phases, the M-site triel atoms participate in covalently bonded anionic chain frameworks with pnictogens (As or Sb), while the A-site species occupy cationic sites around the anionic framework. In Table 2, *Pnma* and *Pbam* phases are indicated in blue and pink, respectively. Notably, the region near $r_A/r_M \approx 0.93$ serves as the structural boundary between the two space groups. Numbers in each cell denote literature references; empty cells indicate 5–2–6 compounds that are, to our knowledge, not yet reported. The two materials discovered in this study are marked with a blue asterisk, whereas theoretically predicted but not synthesized candidates are indicated by a dagger.

This compilation suggests that increasing the size imbalance between A-site and M-site cations tends to favor the *Pnma* structure for 5–2–6 Zintl phases. As shown in Fig. S14b and c, *Pnma* is structurally more complex and less symmetric, and typically associated with a larger unit cell than *Pbam,* with approximately twice the volume[18]. We propose that a large imbalance between the A- and M-site cation size is accommodated by twisting of the polyanionic chains, thereby favoring the *Pnma* space group. As $r_M$ decreases, the M–Pn bond length within the anion-chain framework first shortens[19], and the increased radius mismatch with the A-site cation would otherwise leave the structure inefficiently packed. The structure can then lower its energy by twisting the polyanionic chains, thereby improving the space-filling factor at the expense of an enlarged unit cell. Upon twisting, half of the corner-sharing (M–Pn) tetrahedra of the parent structure becomes edge-sharing, which can reduce the M–M distance to as little as ~58% of the corner-sharing value[20] [00]. Although edge-sharing is electrostatically disfavored according to Pauling's third rule, its adoption in the *Pnma* structure indicates that the gain in packing efficiency outweighs the increased Coulombic repulsion between M-site cations. In contrast, when the size match is more balanced, the more symmetric *Pbam* structure is preferentially stabilized.

This steric perspective also extends to our understanding of the topotactic conversion pathway. As summarized in Table 2, the two space-group regimes respond differently to the parent polytype. In the *Pbam* regime of moderate size imbalance, the conversion is polytype-sensitive: WZ yields the 3–2–4 (*Pnnm*) phase, whereas ZB yields the 5–2–6 (*Pbam*) phase. As observed in the Eu–InAs system, this selectivity appears to hold when the As sublattice remains sufficiently rigid during conversion. In contrast, as the A–M size imbalance increases (e.g., for Ga or Al relative to Eu), the anion sublattice is more likely to relax or reconstruct, allowing the system to bypass parental symmetry constraints and to converge on the $Sr_5Al_2Sb_6$-type (*Pnma*) structural family, regardless of the original polytype (WZ or ZB). This suggests that the governing distinction lies not in the 5–2–6 versus 3–2–4 stoichiometry itself but in the *Pbam* versus *Pnnm* structural competition, plausibly set by the size or volume of the M-centered $Pn_4$ tetrahedra.

While the *Pbam* structure was previously discussed in the context of symmetry-enabled magnetic topologies, the *Pnma* phase may offer distinct advantages for thermoelectric performance. Larger unit cell of the *Pnma* can lead to a reduction in lattice thermal conductivity ($\kappa_L$), thereby enhancing the material's figure of merit ($zT$)[21]. In closely related Zintl phases, *Pnma* structures typically exhibit larger and more complex unit cells than *Pbam*[18]. Recent efforts have further explored deliberate unit-cell expansion in the Zintl systems to improve $zT$[22]. Overall, these results suggest that A–M cation-size imbalance appears to be a steric control parameter for space-group selection in 5–2–6 ternary Zintl phases. Furthermore, the sub-eV band gaps, with values ranging from a few tens of meV in the $Eu_{2n+1}M_2Pn_{2n+2}$ family to a few hundreds of meV in the $Eu_{2n+1}M_nPn_{3n}$ family, offer large tunability, which could be advantageous for photovoltaic and detector applications.

In summary, we have demonstrated the extension of topotaxial mutual-exchange MBE growth to thin films and a variety of parent III-V semiconductors, resulting in the growth of two Zintl compounds, $Eu_5Ga_2As_6$ and $Eu_5Al_2As_6$, that have not been synthesized before by any other method. We have characterized their stoichiometry and structure in complementary SEM, HRTEM combined with EDS, XRD, and DFT calculations. Unlike the previously reported $Eu_3In_2As_4$ (*Pnnm)* and $Eu_5In_2As_6$ (*Pbam*), which both belong to the Zintl family $Eu_{2n+1}M_2Pn_{2n+2}$ of narrow-gap semiconductors, they instead have a distinct crystal structure with *Pnma* point-group symmetry and belong to the Zintl family $Eu_{2n+1}M_nPn_{3n}$ of broad-gap semiconductors. Magnetization measurements reveal a complex magnetic phase diagram, with competing AFM and FM fluctuations that give rise to intricate magnetic structures. This demonstrates the robustness of the topotactic MBE methodology for converting distinct III-V semiconductors into a variety of exotic Eu-based Zintl phases, both as thin films and as NWs. It may, in turn, serve as a platform for a range of technological applications that seamlessly integrate III-V semiconductors and functional Zintl layers.

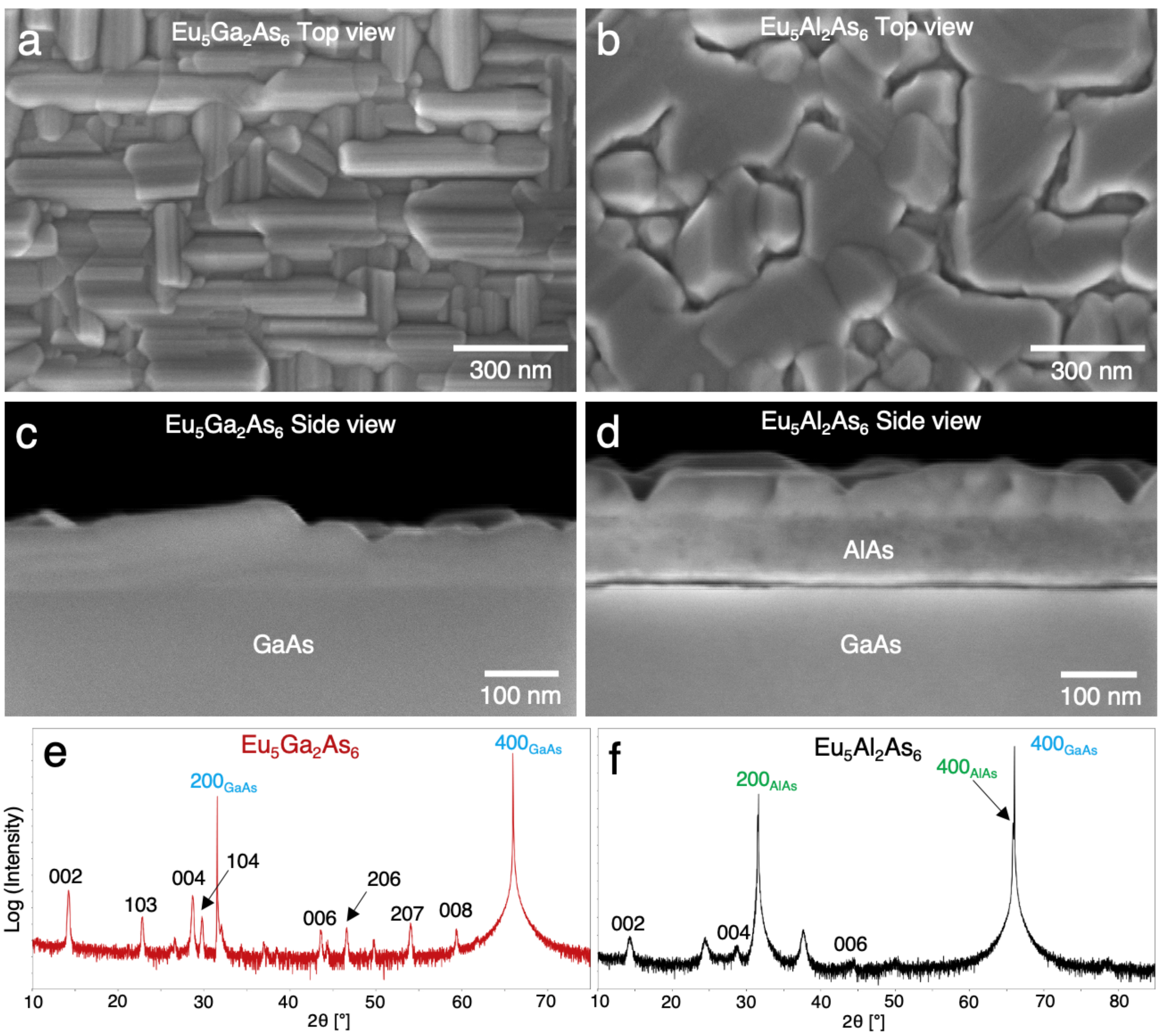

**Figure 1. Mutual exchange growth and XRD patterns of Zintl $Eu_5Ga_2As_6$ and $Eu_5Al_2As_6$ thin films. (a, b)** Top-view SEM images of $Eu_5Ga_2As_6$ and $Eu_5Al_2As_6$ thin films grown on ZB (100) GaAs substrates, respectively. **(c, d)** Side-view SEM images of $Eu_5Ga_2As_6$ and $Eu_5Al_2As_6$ (including AlAs buffer layer) thin films, respectively. **(e)** XRD θ-2θ scan of $Eu_5Ga_2As_6$ thin films grown on GaAs (100) substrates. **(f)** XRD θ-2θ scan of $Eu_5Al_2As_6$ thin films grown on an AlAs buffer layer on GaAs (100) substrate.

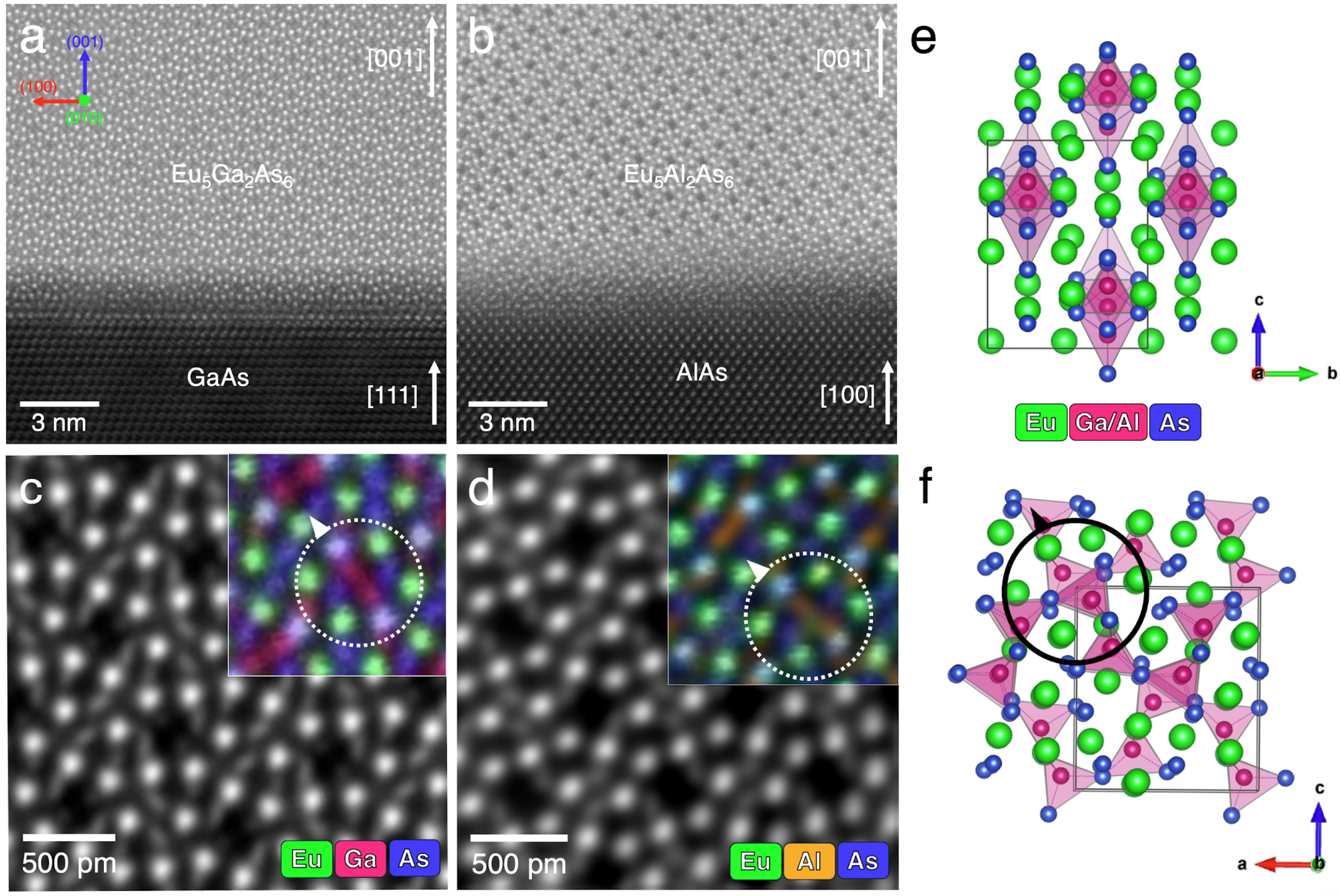

**Figure 2. Topotactic conversion of Zintl $Eu_5Ga_2As_6$ and $Eu_5Al_2As_6$ thin films. (a, b)** High-angle annular dark-field (HAADF) scanning TEM images of the transformation interface between the Zintl thin film and (111)B GaAs substrate/(100) AlAs buffer layer on (100) GaAs substrate, respectively. **(c, d)** Atomic resolution HAADF/STEM images of $Eu_5Ga_2As_6$ and $Eu_5Al_2As_6$, respectively. The inset frames show atomic-scale EDS elemental maps of Eu (green), Ga/Al (magenta/orange), and As (blue), superimposed on the image. A circular marker reveals a superimposed view of chains of juxtaposed $Ga(Al)As_4$ tetrahedra, aligned with the viewing direction in successive layers. **(e, f)** Atomic structure of EuGa(Al)As in the orthorhombic space group *Pnma*, in two viewing directions along [100] and [010]. The experimental data in (c, d) coincide with the [010]-view; the black circle marker in (f) corresponds to the white dashed circle marker in (c, d).

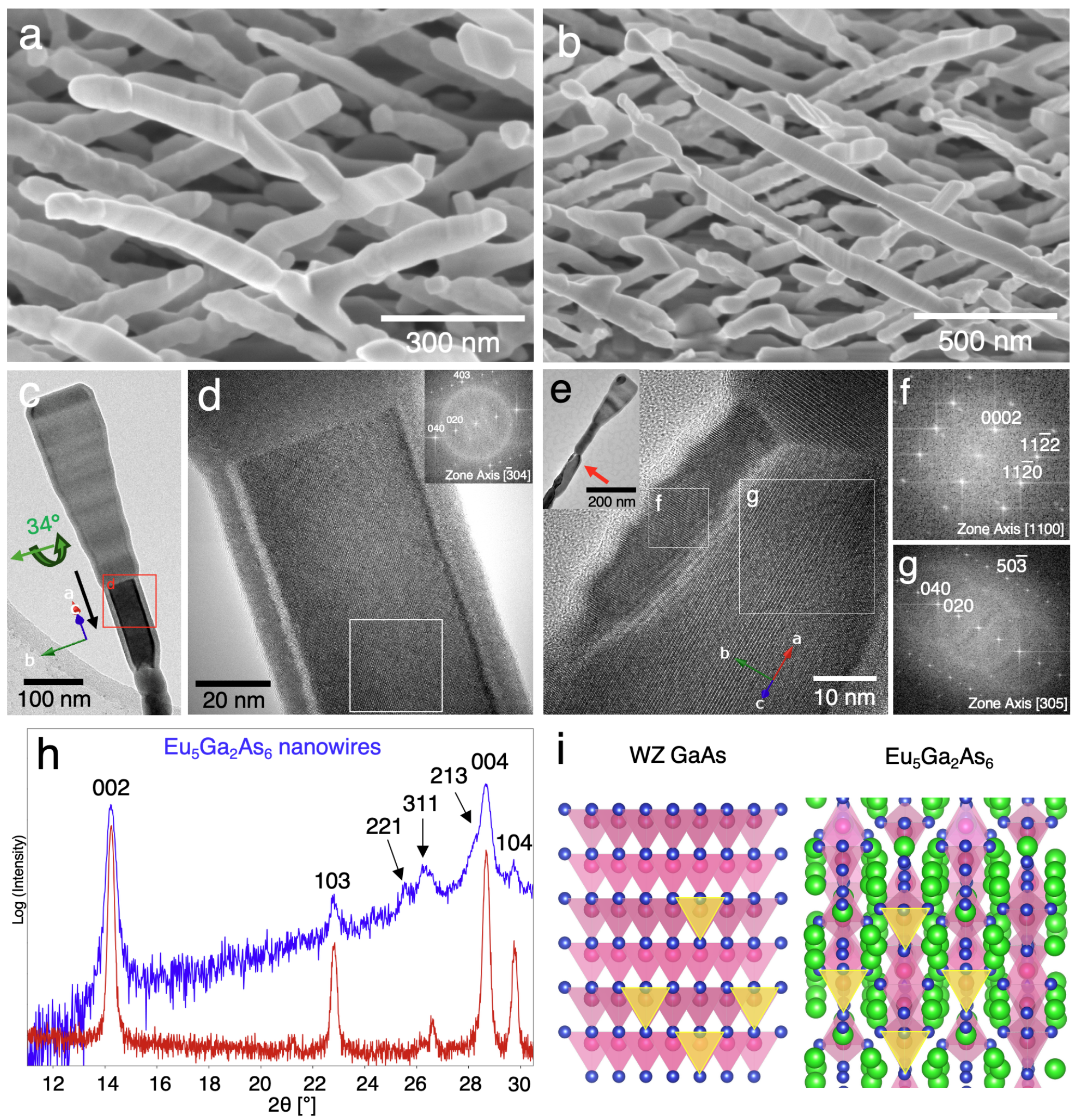

**Figure 3. Zintl $Eu_5Ga_2As_6$ NWs topotactic-exchanged from WZ GaAs core NWs pre-grown on (100) GaAs substrates. (a, b)** Bird's-eye view SEM images of as-grown $Eu_5Ga_2As_6$ NWs. There are two types of Zintl $Eu_5Ga_2As_6$ NWs, showing (a) short-segment and (b) long-segment types. **(c)** TEM image of as-grown $Eu_5Ga_2As_6$ NW fully exchanged from WZ GaAs NW. The middle segment, showing diffraction contrast, has a 34° tilt relative to the adjacent upper segment about the b-axis [Fig S07 in SI]. The arrow compass of the unit cell signifies the $[\underline{3}04]$ zone axis of the middle segment. **(d)** An HRTEM image of the area marked by a red square in (c) and an FFT pattern (inset) of the area marked by a white square in (d). **(e)** An HRTEM image of the interface between WZ GaAs and Zintl regions (inset: low-magnification TEM image with a red arrow indicating the corresponding area). **(f, g)** FFT patterns of the regions outlined by white squares in (e), corresponding to WZ GaAs along the $[1\underline{1}00]$ axis (f) and Zintl $Eu_5Ga_2As_6$ along the [305] axis (g). **(h)** XRD θ-2θ scan of as-grown $Eu_5Ga_2As_6$ NWs on GaAs (100) substrates (blue line); for comparison, data from Fig. 1e is overlaid (red line). **(i)** Atomic model of the topotactic transformation at the interface between the WZ GaAs (f) and Zintl $Eu_5Ga_2As_6$ regions (g).

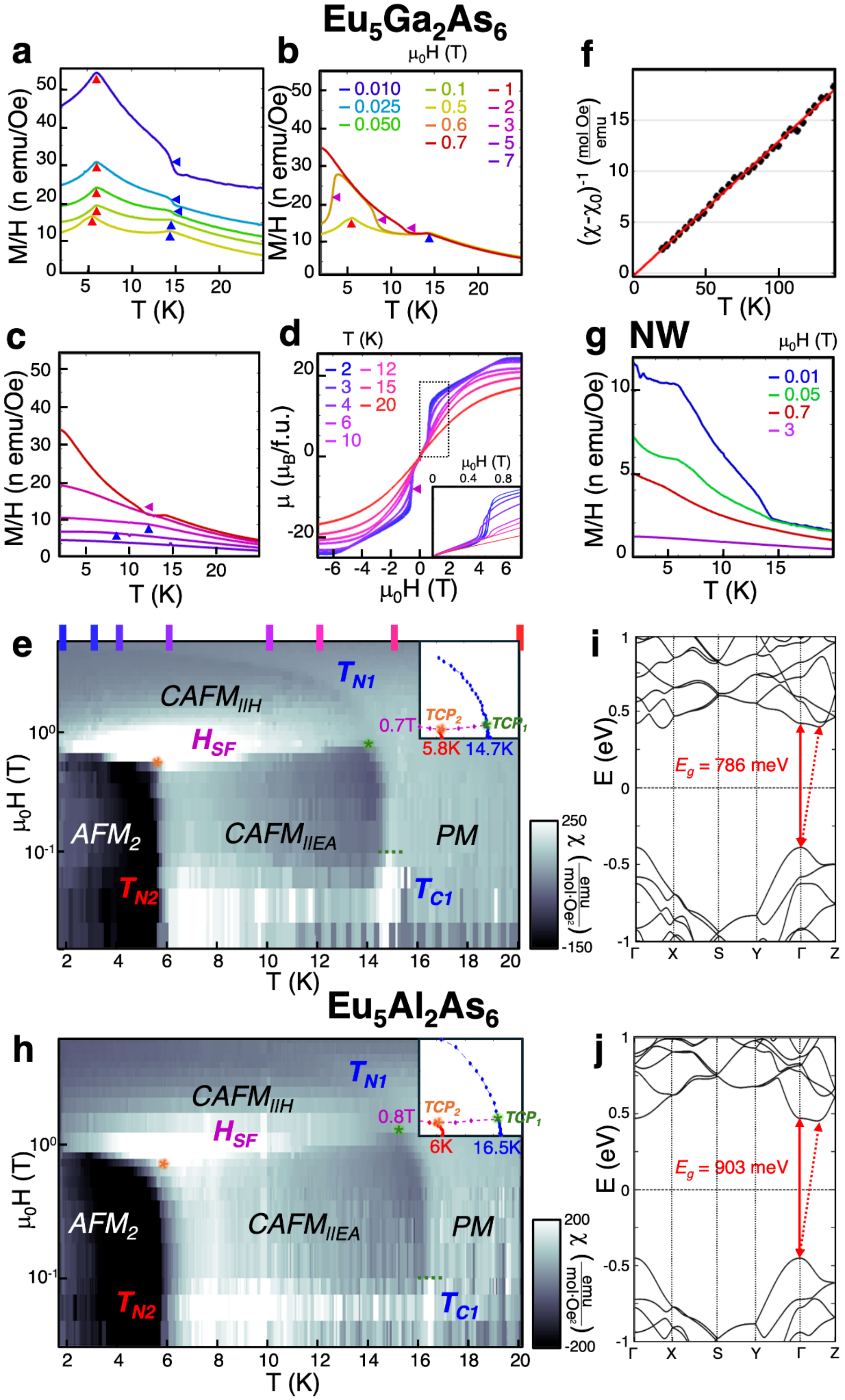

Eu5Ga2As6
a
b
c
d
e
f
g
h
i
j
NW
Eu5Al2As6
M/H (n emu/Oe)
T (K)
μ0H (T)
μ (μB/f.u.)
CAFMIIH
CAFMIIEA
HSF
AFM2
PM
TN1
TN2
TC1
TCP1
TCP2
0.7T
5.8K
14.7K
0.8T
6K
16.5K
Eg = 786 meV
Eg = 903 meV
E (eV)
Γ
X
S
Y
Z

**Figure 4. Magnetic phase diagram and ab initio electronic band structures of Zintl $Eu_5Ga_2As_6$ and $Eu_5Al_2As_6$. (a)** Magnetization of $Eu_5Ga_2As_6$ thin film measured in MPMS along temperature sweeps at varying magnetic fields. Triangles indicate transitions. The inset shows a magnetization measurement over a broad temperature range with a Curie-Weiss linear fit (red line). **(b)** Magnetic moment $\chi$=dM/dT calculated from temperature sweeps. The inset shows the extracted transition lines. **(c)** Magnetization measured along field sweeps at different temperatures. **(d)** Same as b, but for $Eu_5Al_2As_6$ **(e)** Magnetization of $Eu_5Ga_2As_6$ nanowires harvested and dispersed over Si/$SiO_2$ substrate measured along temperature sweeps at different applied magnetic fields. **(f,g)** Band structure of $Eu_5Ga_2As_6$ and $Eu_5Al_2As_6$ of the paramagnetic phase calculated in DFT.

**Table 1. Lattice constants of $Eu_5Ga_2As_6$ from XRD, HAADF-STEM, and DFT.**

| Method | a (Å) | b (Å) | c (Å) |
|---|---|---|---|
| **XRD (thin films + NWs)** | 11.374 | 9.4177 ± 0.0094 | 12.437 ± 0.0004 |
| **HAADF-STEM (HCFI)** | 11.0410[1] | 9.3433[2] | 12.1579[1]/12.3548[2] |
| **DFT (0 K)** | 11.4833 | 9.7404 | 12.6054 |

[1] measured along [010] zone axis
[2] measured along [100] zone axis

**Table 2. Map of Zintl $A_5M_2Pn_6$ space group *Pnma* vs. *Pbam* (A = Yb, Ca, Eu, Sr, Ba, M = Al, Ga, In, Tl, Pn = As, Sb).** Summary of reported $A_5M_2Pn_6$ Zintl compounds classified by space group, color-coded in light orange and light blue for *Pnma* and *Pbam*, respectively. Each cell lists the literature references for compounds known experimentally or theoretically; blank cells indicate no reported references. The number in each cell represents the radius ratio ($r_A/r_M$) of the magnetic ion (A) to the Group 13 element (M); ratios based on Shannon ionic radii[23] for A and Cordero covalent radii[24] for M. (* what we found, †theoretically exist, ‡$Ca_5Al_2As_6$ was mentioned as a space group, *Pbam*, in Ref. 25, but no structural data has been reported to date.

| M | Al | | | Ga | | | | In | | | | Tl | | |
|---|---|---|---|---|---|---|---|---|---|---|---|---|---|---|
| Pn \ A | As | ratio | Sb | As (WZ) | As (ZB) | ratio | Sb | As (WZ) | As (ZB) | ratio | Sb | As | ratio | Sb |
| Yb | - | 1.07 | [25] | - | - | 1.08 | [26] | - | - | 1.26 | [27] | - | 1.28 | [7†] |
| Ca | [28‡] | 1.06 | [29–31] | - | [19,32] | 1.07 | [33–35] | - | - | 1.25 | [19,33] | - | 1.27 | - |
| Eu | * | 0.92 | [36] | * | * | 0.93 | [7†,37†] | [4] | [5,38,39] | 1.08 | [40] | - | 1.11 | [7†] |
| Sr | - | 0.92 | [28,29,41] | - | - | 0.92 | - | - | [38] | 1.08 | [33] | - | 1.10 | - |
| Ba | - | 0.81 | - | - | - | 0.82 | - | | - | 0.95 | [28] | - | 0.97 | - |

$r_A/r_M$ ●:• → ***Pnma***  $r_A/r_M$ ●:● → ***Pbam***  $0.92 \lesssim r_A/r_M \lesssim 0.95$ → ***Pnma*** or ***Pbam***  ***Pnnm***

# Methods

## Zintl-phase films and NWs growth

Zintl $Eu_5Ga_2As_6$ films were topotaxially grown on GaAs (100) and (111)B substrates in a high purity MBE (RIBER-32) system. The GaAs substrate, regardless of orientation, underwent an oxide blow-off process without As overpressure in a separate chamber attached to the MBE. For the oxide blow-off, the temperature is ramped up to ∼ 660 ºC at a rate of 10 ºC $min^{-1}$, remaining for 5 min at the peak temperature and ramped down at a rate of 20 ˚C $min^{-1}$ to ∼ 100 ºC. The sample is then moved to the growth chamber and held at 370 ºC until As overpressure builds up to ∼ $2 \times 10^{-6}$ Torr. To initiate the topotaxial cation exchange, the Eu shutter is opened, and the substrate temperature is ramped up to 570 ºC at a rate of 10 ˚C $min^{-1}$. The topotaxial exchange growth is maintained for 3 hours. The temperature and flux of Eu are 450 ºC and $3.8\times10^{-8}$ Torr, respectively. For $Eu_5Al_2As_6$ films, the AlAs buffer layer were first grown on a GaAs (100) substrate for 1 hour at a substrate temperature of 570 ºC, with the Al and As cells at 1000 ºC ($2.2\times10^{-8}$ Torr) and 235 ºC ($4.0\times10^{-6}$ Torr), respectively. The subsequent topotaxial exchange was carried out for 2 hours under the same conditions as for the $Eu_5Ga_2As_6$ films.

Zintl $Eu_5Ga_2As_6$ NWs were topotaxially grown on the surface of WZ GaAs NWs. As a preliminary step, reclining and vertical WZ GaAs NWs were grown by MBE on (100) or (110), and (111)B GaAs substrates, respectively, using the gold-assisted vapor-liquid-solid (VLS) technique as described in the previous work[4,5]. To initiate the topotaxial cation exchange, the Eu shutter was opened while the Ga cell was cold and its shutter closed with the As kept. A 15-minute pause for adjusting the cell temperatures followed the GaAs NWs growth. Immediately after opening of the Eu shutter the substrate temperature was ramped to a temperature of 570 ºC at a rate of 10 ºC $min^{-1}$.

The topotaxial exchange growth was maintained for 2.5 hours. The temperature (and flux) of Eu and As were 450 ºC ($3.5\times10^{-8}$ Torr) and 220 ºC ($2.0\times10^{-6}$ Torr), respectively.

**Microscopy**

The Zintl $Eu_5Ga_2As_4$ films/NWs and $Eu_5Al_2As_4$ films were characterized by field emission scanning electron microscopy (FE-SEM, Zeiss Supra-55, 3 kV, working distance ~4 mm), transmission electron microscopy (TEM, Thermo Fisher Scientific Talos F200X, 200 kV). Energy dispersive spectroscopy (EDS) composition data and mapping images were obtained by TEM with an attached detector that is identical to the one in the scanning transmission electron microscopy (STEM).

High-resolution STEM images and analytical EDS maps were acquired in a double aberration-corrected Themis-Z microscope (Thermo Fisher Scientific Electron Microscopy Solutions, Hillsboro, USA, (TFS)) at an accelerating voltage of 200 kV. STEM images were recorded with a Fischione Model 3000 detector and a TFS BF detector. EDS hyperspectral data were obtained with a Super-X SDD detector and quantified with the Velox software (TFS) through background subtraction and spectrum deconvolution. STEM images were obtained with an electron probe with a convergence angle of 21 mrad and a primary beam current of less than 50 pA, the EDS maps were recorded at a beam current of 200 pA. To examine a cross-sectional structure of the thin films, a lamella preparation in a focused ion beam (FIB) was performed using a thick lift-out procedure (Helios 600 FIB/SEM Dual Beam Microscope, Thermo Fisher Scientific). The lamella (~ 50 nm thick) was placed on a TEM grid.

**Magnetization measurements**

Magnetization measurements were acquired using a commercial Quantum Design SQUID Magnetic Properties Measurement System (MPMS-3), equipped with a 1.6 K $^4$He cryostat and a 7 T magnet. The samples were mounted in a non-magnetic straw and loaded in the equipment.

**First-principles calculations**

First-principles calculations are performed using the Vienna Ab initio Simulation Package (VASP) using the projector-augmented-wave method[42]. The exchange-correlation functional is treated with the generalized gradient approximation (GGA) using the Perdew-Burke-Ernzerhof (PBE) parametrization[43]. To treat the correlation effect of localized 4f electrons of Eu, the density functional theory (DFT) + $U$ method was employed, with $U_{eff} = 7$ eV[44]. The kinetic-energy cutoff of the plane-wave basis set was 450 eV. Brillouin-zone integration was performed by using a 5 × 5 × 5 Γ-centred k-point mesh.

**Data availability**

**Acknowledgements**

M.S.S. acknowledges support from the National Research Foundation of Korea (NRF) grant funded by the Korea government (MSIT) (RS-2026-25480768) and by Global – Learning & Academic research institution for Master's·PhD students, and Postdocs (LAMP) Program of the National Research Foundation of Korea (NRF) grant funded by the Ministry of Education (RS-2025-25440216). L.H. acknowledges support from the Irving and Cherna Moskowitz Center for Nano and Bio-Imaging at the Weizmann Institute of Science. H.S. is an incumbent of the Henry and Gertrude F. Rothschild Research Fellow Chair. B.Y. acknowledges financial support from the European Research Council (ERC Consolidator Grant ʹNonlinearTopoʹ, no. 815869) and the ISF— Personal Research Grant (no. 2932/21). H.B. and H.S. acknowledge support from the European Research Council (ERC-PoC TopoTapered—101067680) and the Israel Science Foundation (grant 1152/23).

**Author contributions**

M.S.S., H.S. and H.B. conceptualized this research. H.S. and M.S.S. designed and carried out the MBE growths. M.S.S. and L.H. conducted the electron microscopy imaging with EDS analysis and modelling of the interfacial crystal structure. Y.Z. and B.Y. performed the DFT calculations. J.S., E.F. M.H. and A.G. performed the magnetic measurements and analysed the data. A-.E.K. carried out the XRD measurements and interpreted the data. M.S.S., L.H., H.S. and H.B. prepared the figures in the manuscript and Supplementary Information. All authors contributed to the discussion and manuscript preparation.

**Competing interests**

The other authors declare no competing interests.

## SUPPLEMENTARY INFORMATION

# Discovery of New Zintl Films and Nanowires Grown by Topotaxy Conversion of III-V Semiconductors

Man Suk Song,[1,2] Lothar Houben,[3] Jean Souza,[1] Edanel Fishbein[1], Moshe Haim[1], Ambikesh Gupta,[1] Yufei Zhao,[1,4] Anna-Eden Kossoy,[3] Binghai Yan, [1,4] Haim Beidenkopf,[1]* and Hadas Shtrikman[1]*

[1]Department of Condensed Matter Physics, Weizmann Institute of Science, Rehovot 7610001, Israel

[2]Department of Semiconductor, Dong-A University, Busan 49315, Republic of Korea

[3]Department of Chemical Research Support, Weizmann Institute of Science, Rehovot, 7610001, Israel

[4]Department of Physics, The Pennsylvania State University, University Park, PA, USA

*E-mail: haim.beidenkopf@weizmann.ac.il, hadas.shtrikman@weizmann.ac.il

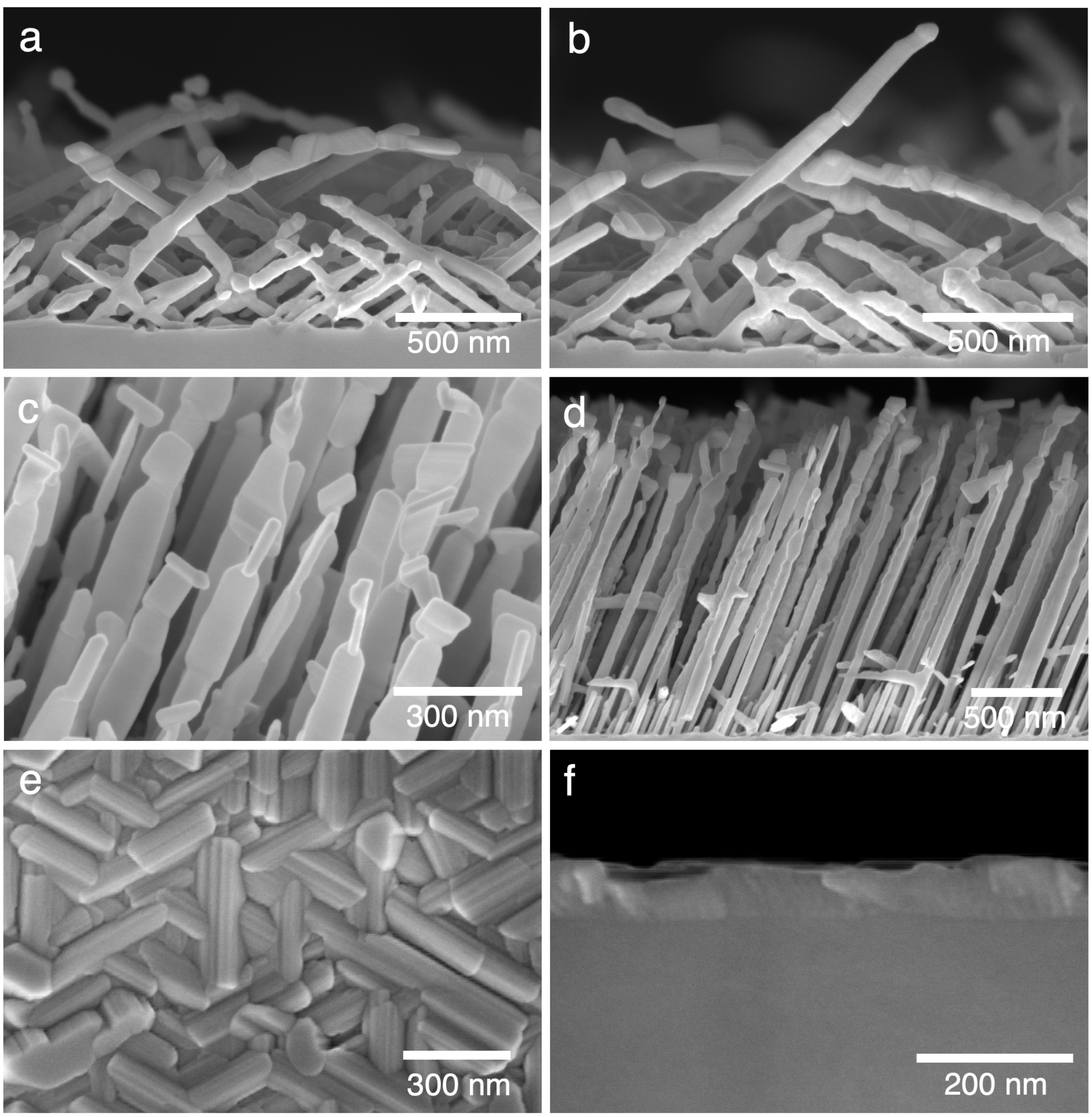

**Figure S01. Mutual exchange growth of $Eu_5Ga_2As_6$ thin films and NWs. (a, b)** Side-view SEM images of as-grown $Eu_5Ga_2As_6$ NWs on ZB (100) GaAs substrates. There are two types of $Eu_5Ga_2As_6$ NWs, showing short-segment (a) and long-segment (b) types. **(c, d)** Bird's eye-view and side-view SEM images of as-grown $Eu_5Ga_2As_6$ NWs on ZB (110) GaAs substrates, respectively. **(e, f)** Top-view and side-view SEM images of $Eu_5Ga_2As_6$ thin films on ZB (111)B GaAs substrates, respectively.

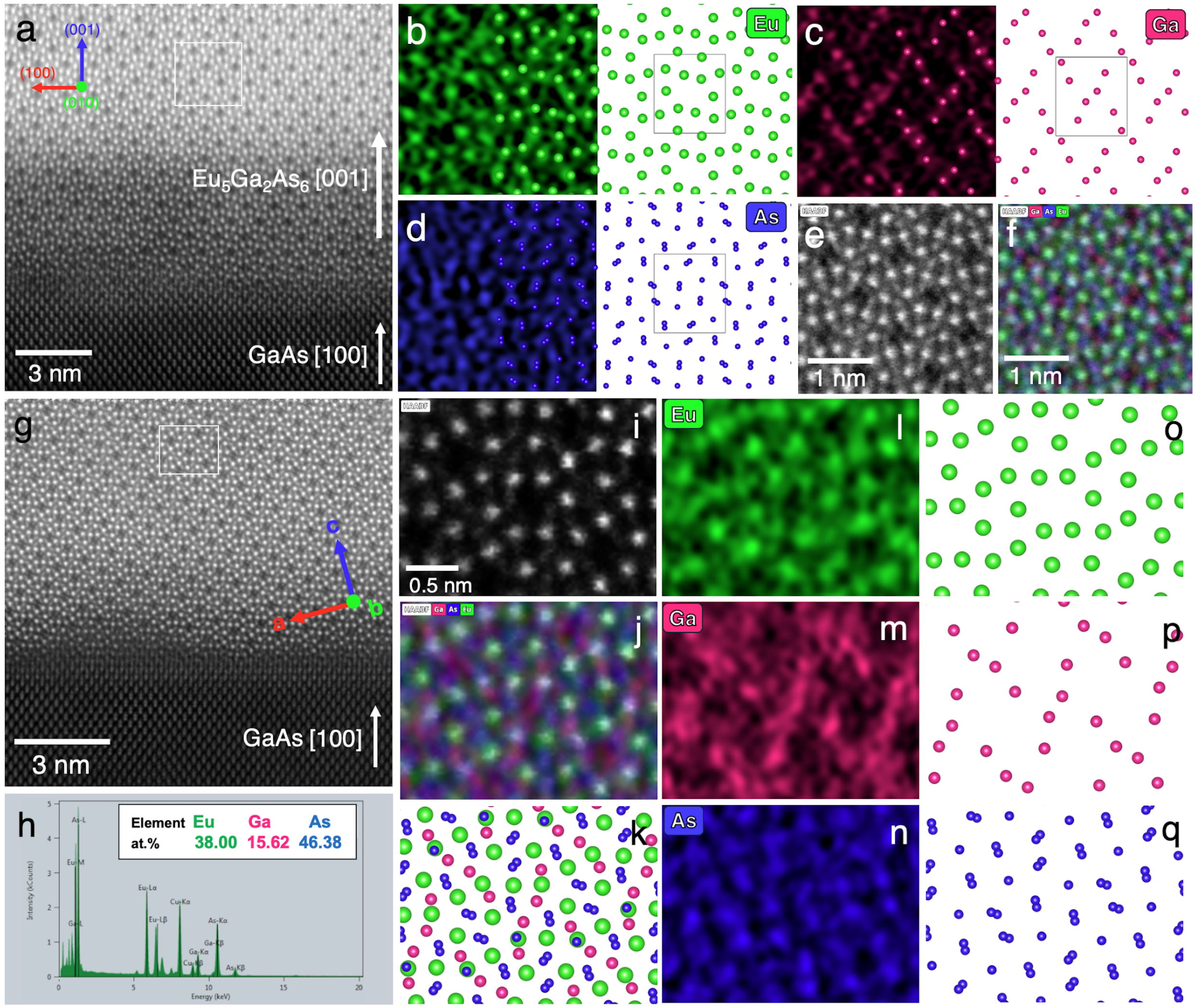

**Figure S02. STEM-HAADF and EDS analyses of $Eu_5Ga_2As_6$ thin films on ZB GaAs (100) substrate. (a)** STEM-HAADF image of the interface between the $Eu_5Ga_2As_6$ thin film, which grows along the c axis, and the GaAs (100) substrate. **(b–d)** EDS elemental maps of Eu (green), Ga (magenta), and As(blue), respectively; an atomic model is continuously overlaid on the right-hand side. The outlined rectangle indicates a 1×1 unit cell of $Eu_5Ga_2As_6$ viewed along the [010] zone axis. **(e, f)** High-resolution HAADF images of the $Eu_5Ga_2As_6$ film and the corresponding EDS elemental maps of Eu, Ga, and As, respectively. **(g)** STEM-HAADF image of the interface between the $Eu_5Ga_2As_6$ thin film and the (100) GaAs substrate. The $Eu_5Ga_2As_6$ film grows along the $[\bar{1}03]$ direction. **(h)** EDS spectrum extracted from the region outlined by the white square in (g), with the inset showing the quantitative analysis. **(i, j)** High-resolution STEM-HAADF image of the $Eu_5Ga_2As_6$ thin film and the corresponding EDS elemental maps of Eu, Ga, and As, respectively. **(k)** Atomic model corresponding to (i) and (j). **(l–n)** EDS elemental maps of Eu, Ga, and As, respectively. **(o–q)** Individual atomic models of Eu, Ga, and As corresponding to (l), (m), and (n), respectively.

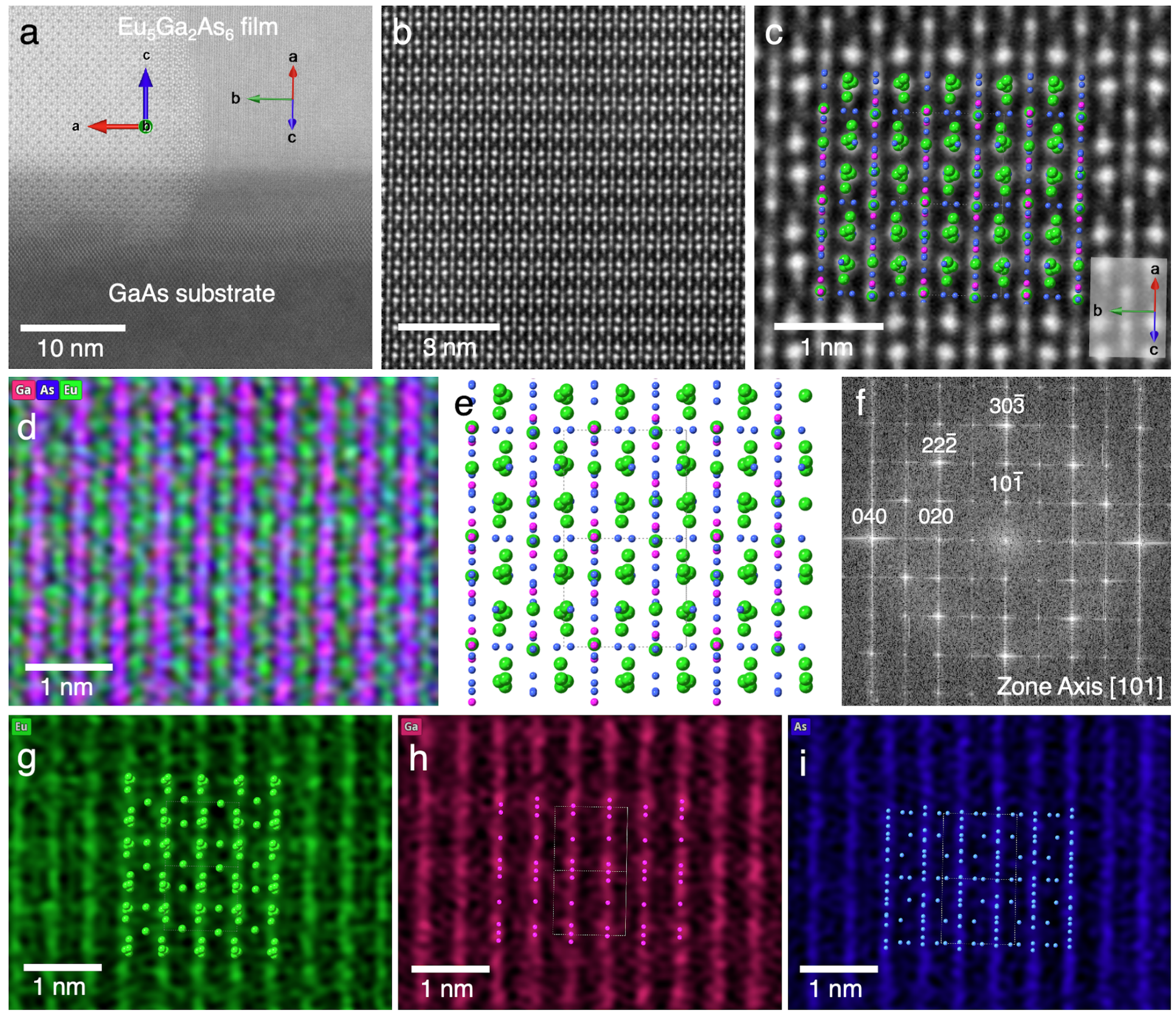

**Figure S03. STEM-HAADF and EDS analyses of $Eu_5Ga_2As_6$ thin films on ZB (111)B GaAs substrates. (a)** STEM-HAADF image of the interface between the $Eu_5Ga_2As_6$ thin films and the (111)B GaAs substrates. The $Eu_5Ga_2As_6$ thin film on the left grows along the [001] direction, parallel to the surface normal of the (111)B substrate, while the film on the right grows along the [10-1] direction, likely originated from a different initial interface. **(b)** High-resolution STEM-HAADF image of the $Eu_5Ga_2As_6$ thin film along the $[10\underline{1}]$ direction. **(c)** Enlarged image of (b) with the atomic structure of $Eu_5Ga_2As_6$ overlaid. The gray dashed lines denote the unit cell. **(d, e)** EDS maps of Eu (green), Ga (magenta), and As (blue), and atomic model, respectively, corresponding to (c). **(f)** FFT pattern from b, showing the [101] zone axis. (g–i) EDS elemental maps of Eu, Ga, and As, respectively. Each atomic model is overlaid, and the gray dashed lines indicate the unit cell.

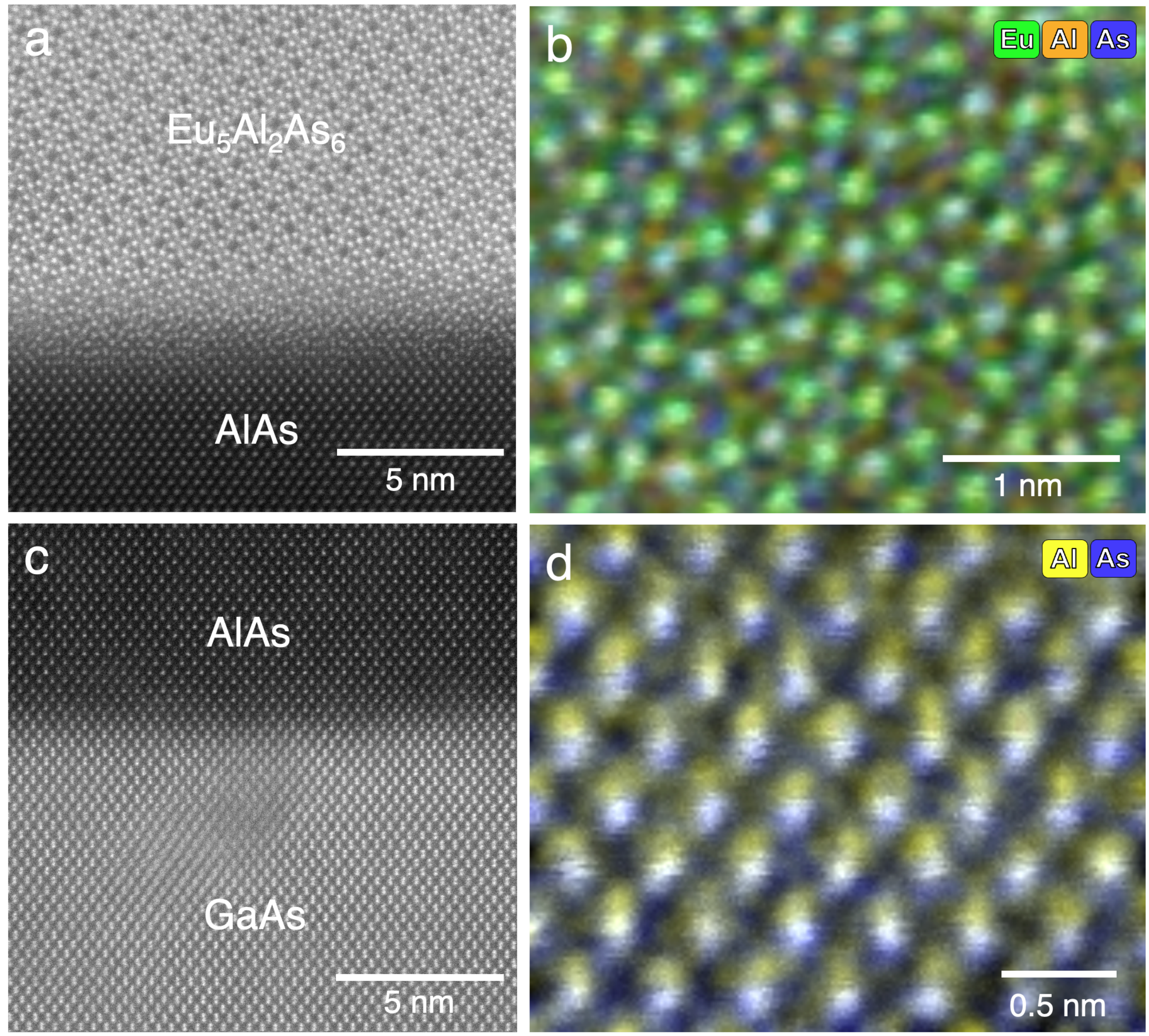

**Figure S04. STEM-HAADF and EDS analyses of $Eu_5Al_2As_6$ thin films grown on an AlAs buffer layer on (001) GaAs substrates. (a)** STEM-HAADF image of the interface between the $Eu_5Ga_2As_6$ thin film and the AlAs buffer layer. **(b)** EDS maps of Eu (green), Al (orange), and As (blue). **(c)** STEM-HAADF image of the epitaxial interface between the AlAs buffer layer and the (001) GaAs substrates. **(d)** EDS maps of Al (yellow) and As (blue).

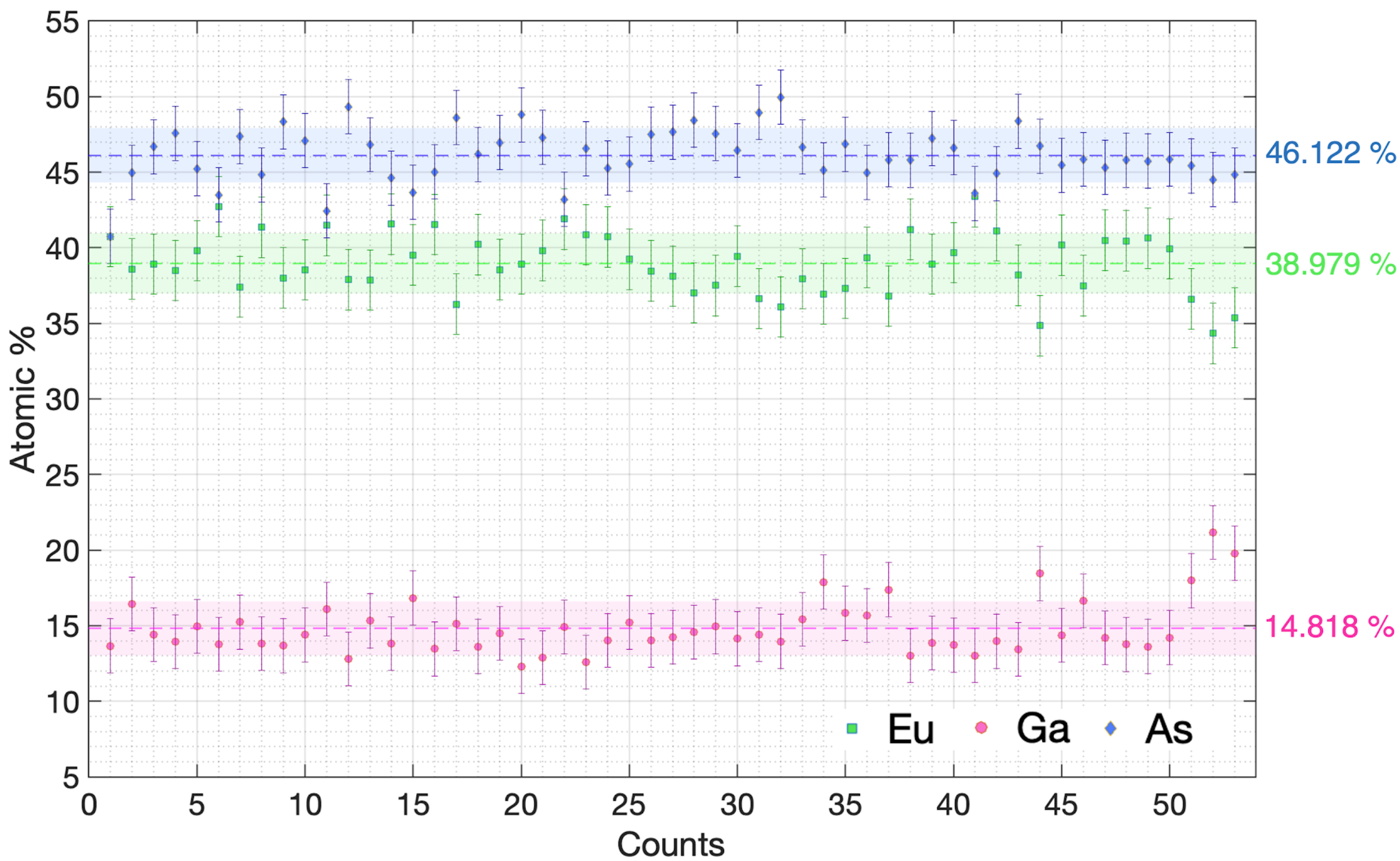

**Figure S05. STEM-EDS composition statistics of $Eu_5Ga_2As_6$ NWs.** Atomic percentages (circular markers) from n = 53 measurements along multiple TEM-identified NWs are shown for Eu (green), Ga (magenta), and As (blue). Dashed lines indicate the mean; shaded bands show mean ± 1σ; vertical ticks are error bars (global standard deviation [SD] applied to each point). The averaged composition is Eu 38.97 ± 2.00 atomic %, Ga 14.82 ± 1.79 atomic %, As 46.12 ± 1.80 atomic % (mean ± SD), in excellent agreement with the nominal Eu:Ga:As = 5:2:6 stoichiometry (theoretical 38.46:15.38:46.15 atomic %), with deviations ≤ 0.6 atomic % for each element.

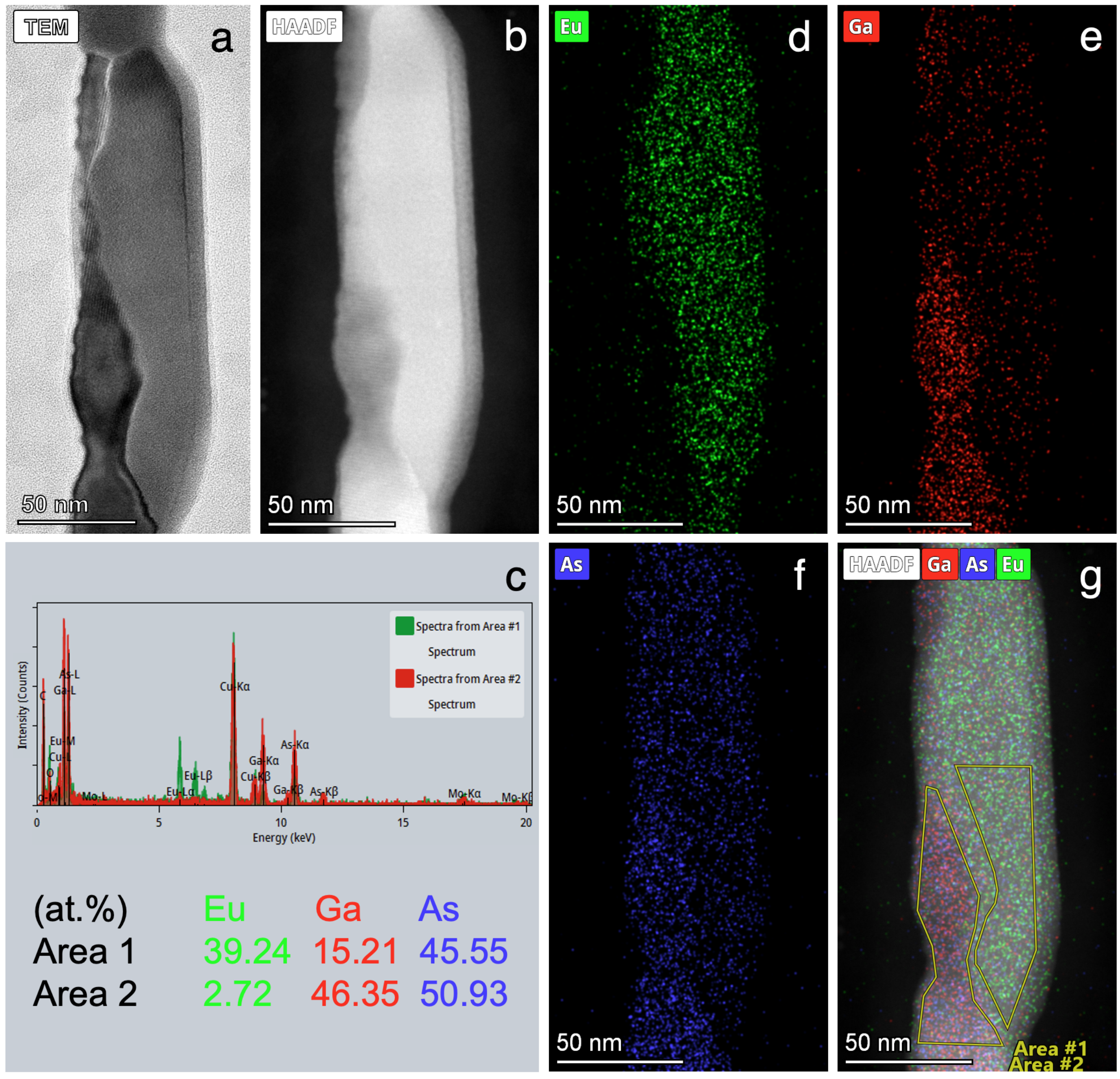

**Figure S06. Elemental composition of orthorhombic $Eu_5Ga_2As_6$ and wurtzite GaAs region in the NW. (a, b)** TEM and HAADF images of the $Eu_5Ga_2As_6$ and the GaAs regions selected for EDS measurements. **(c)** Cumulated EDS spectra extracted from the two regions, Area #1 and Area #2, indicated in g. **(d–f)** EDS elemental maps showing the distributions of Eu (green), Ga (red), and As (blue), respectively. **(g)** Composite map combining HAADF with Eu, Ga, and As elemental maps.

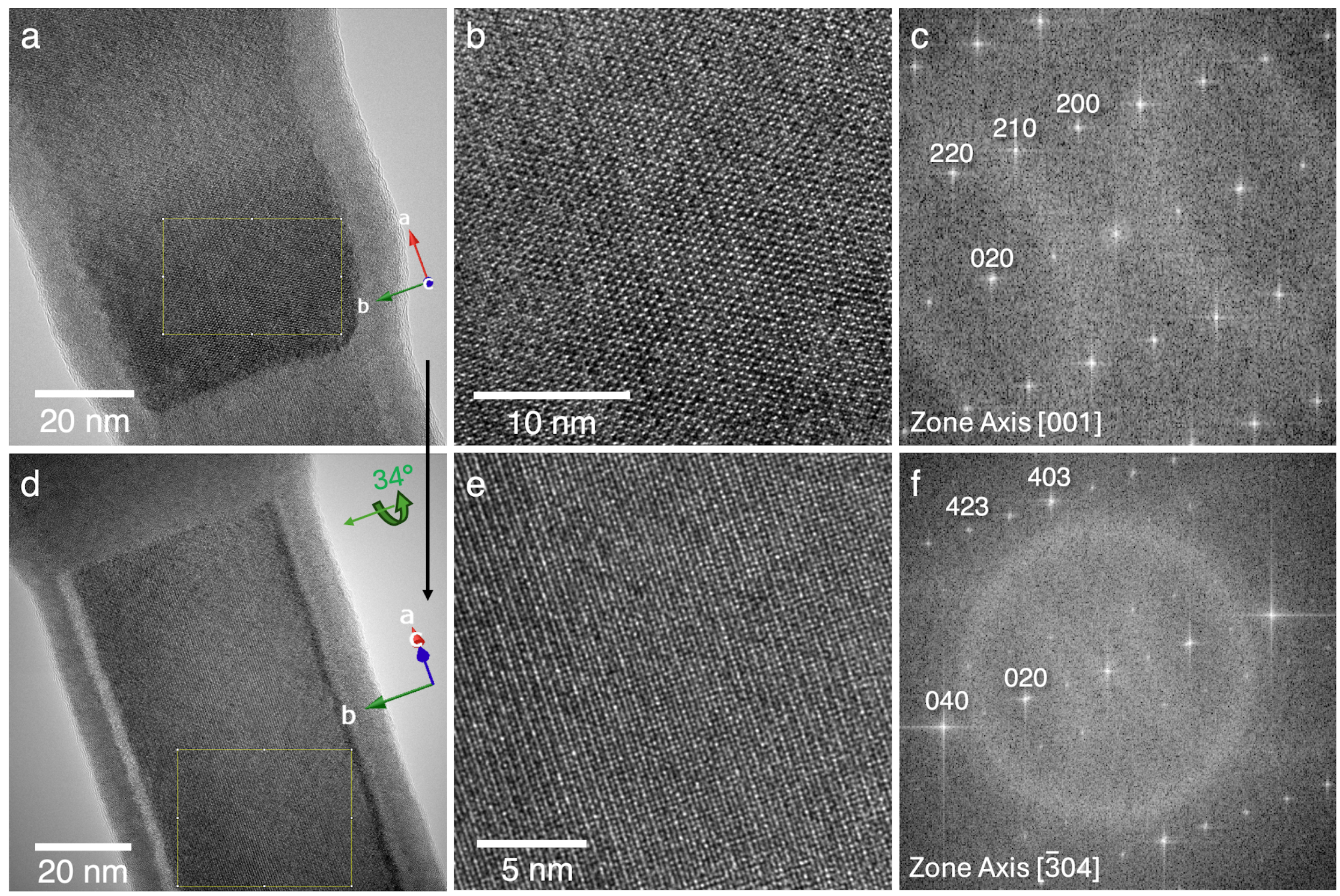

**Figure S07. HRTEM and FFT analyses of $Eu_5Ga_2As_6$ NWs (long segment type)**. **(a, d)** TEM images of the $Eu_5Ga_2As_6$ NWs, identical to those shown in Figure 03(b) and 03(d) of the main text, respectively. These images were obtained in the same NW but at a 34° tilt to align the zone axes of individual $Eu_5Ga_2As_6$ segments; arrows indicate the zone axes within the yellow rectangles. **(b, d)** HRTEM images corresponding to the areas marked by yellow rectangles in **(a)** and **(d)**, respectively. **(c, f)** FFT patterns of the regions in (b) and (e), indicating the [001] and [$\underline{3}$04] zone axes of $Eu_5Ga_2As_6$, respectively.

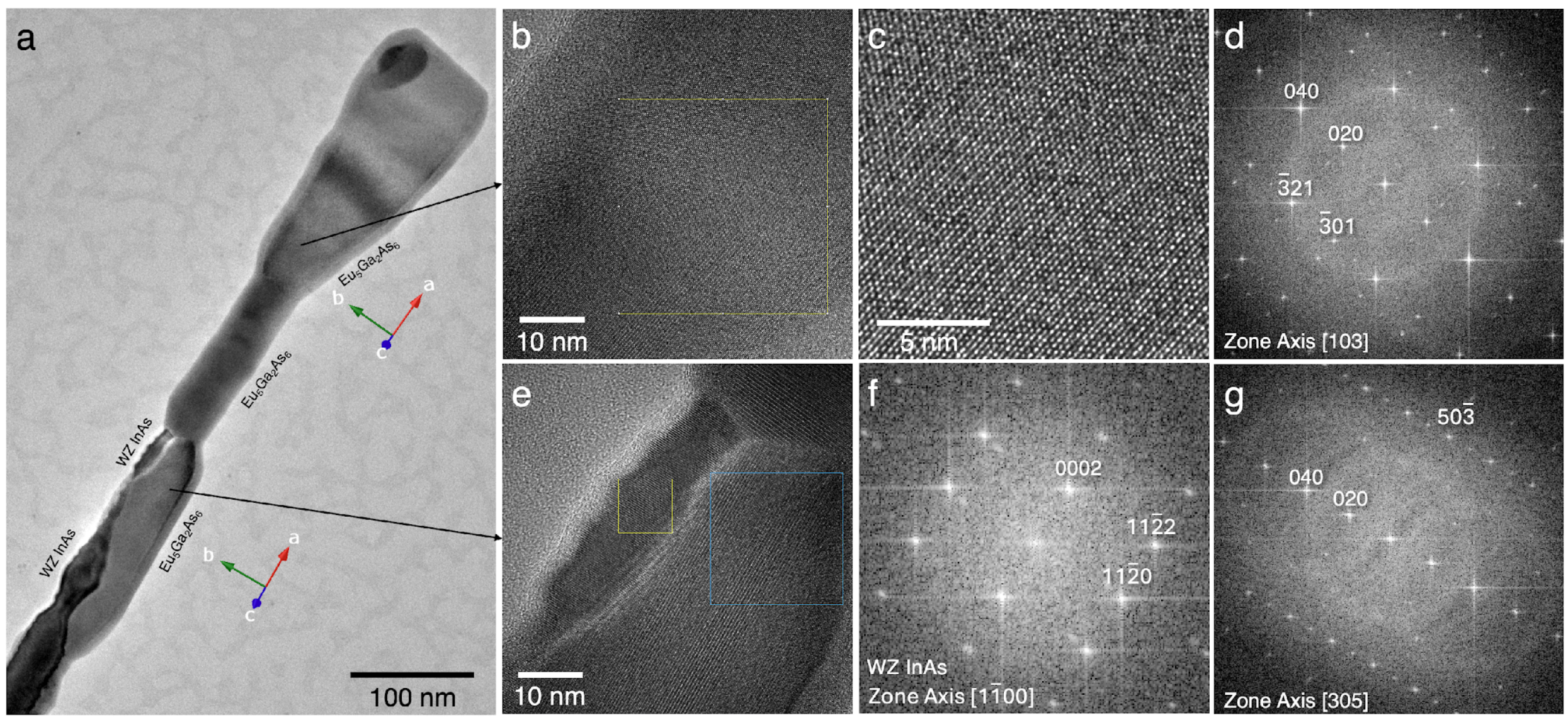

**Figure S08. HRTEM and FFT analyses of $Eu_5Ga_2As_6$ NWs (long segment type). (a)** TEM image of another long segment $Eu_5Ga_2As_6$ NW; arrows indicate the [103] and [305] zone axes of each segment, which are nearly parallel. **(b, e)** HRTEM images of the $Eu_5Ga_2As_6$ segment in a completely topotactic exchanged region and in the interface between WZ InAs and $Eu_5Ga_2As_6$, respectively. **(c, d)** Enlarged HRTEM image and FFT pattern of the area indicated by the yellow square in (b), respectively. **(f, g)** FFT patterns of the area marked by the yellow and blue squares in (e), respectively.

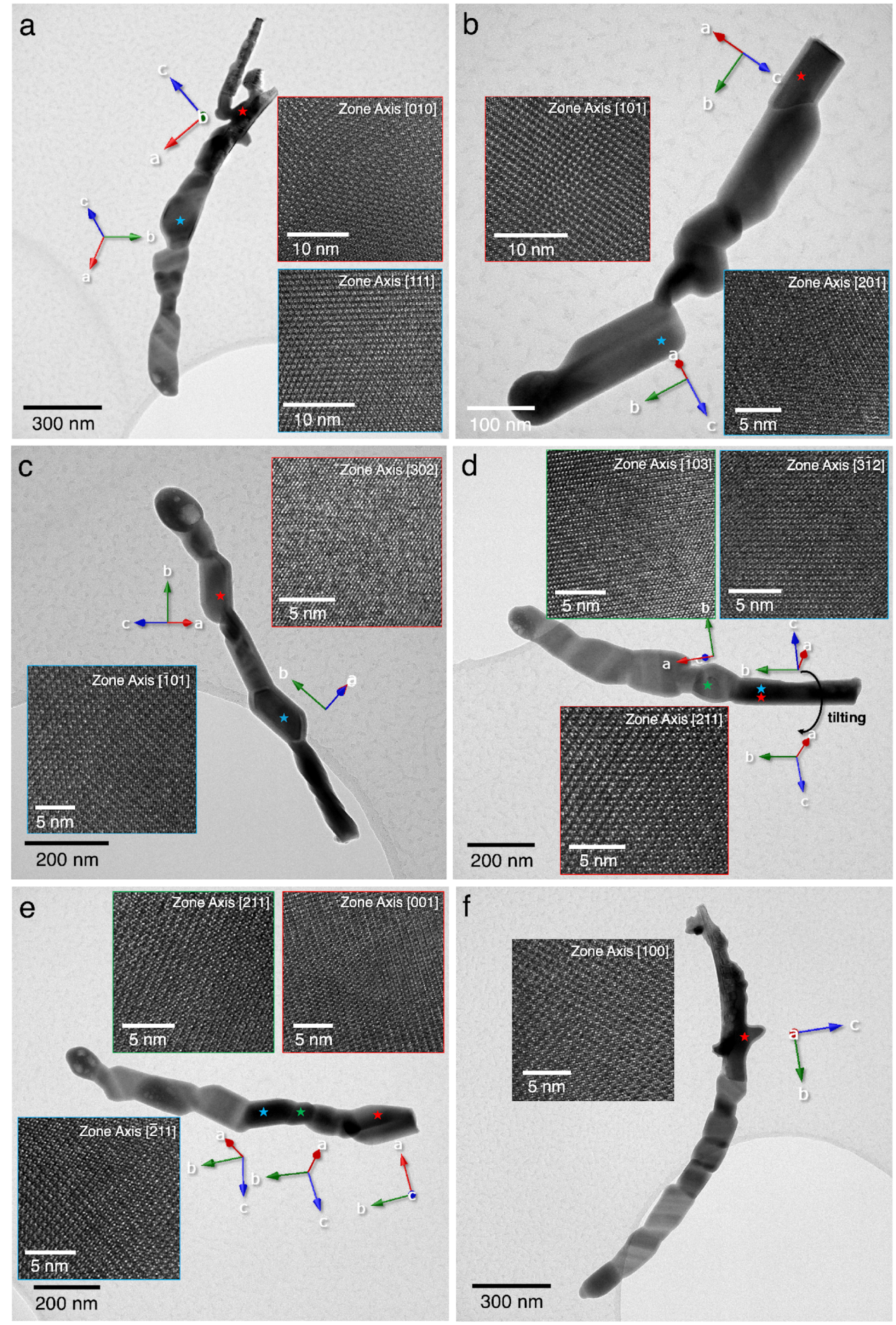

**Figure S09. HRTEM and FFT analyses of $Eu_5Ga_2As_6$ NWs (short segment type). (a–f)** TEM images of short segment $Eu_5Ga_2As_6$ NWs. Insets show HRTEM images corresponding to the areas marked by red, blue, and green stars. Arrows indicate the zone axes of the respective areas.

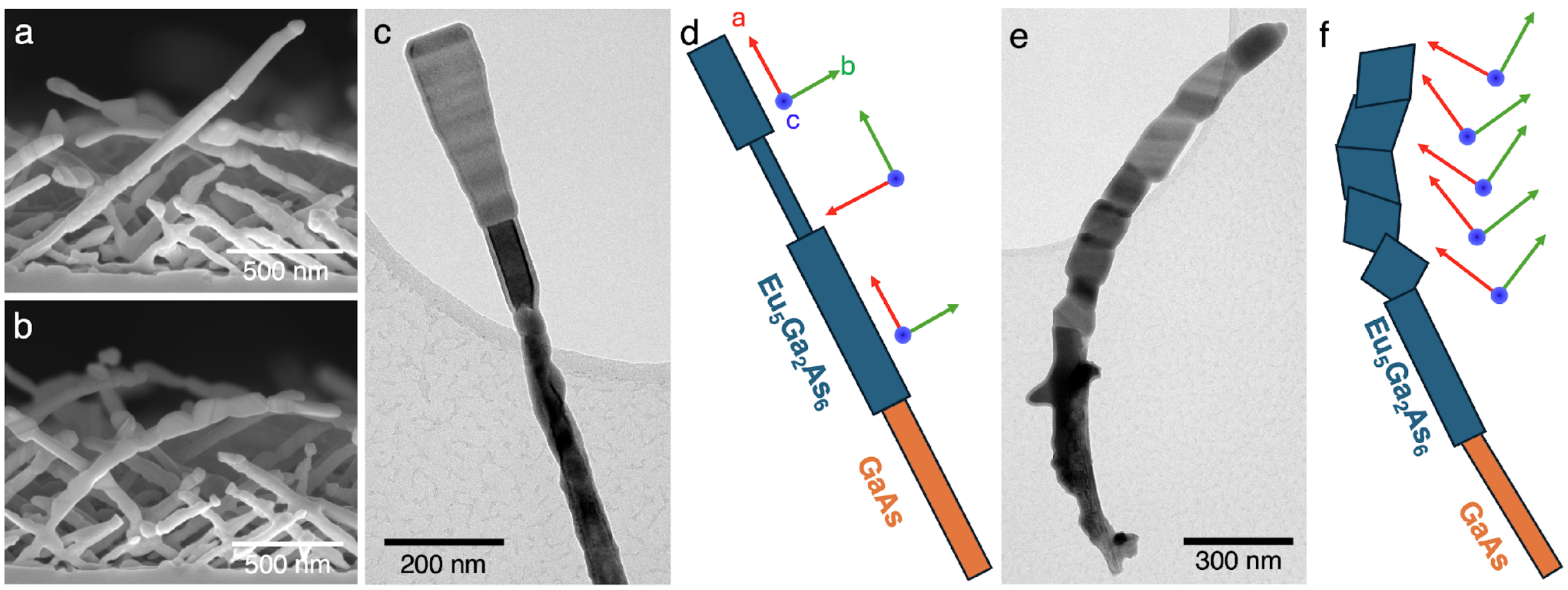

**Figure S10. Different types of $Eu_5Ga_2As_6$ NWs: long and short segments. (a, b)** Side-view SEM images showing representative $Eu_5Ga_2As_6$ NWs with long and short segments, respectively. **(c, d)** TEM and schematic images of a long-segment type $Eu_5Ga_2As_6$ NW, respectively. **(e, f)** TEM and schematic images of a short-segment type $Eu_5Ga_2As_6$ NW, respectively. In (d) and (f), arrows indicate zone axes of the adjacent $Eu_5Ga_2As_6$ segments.

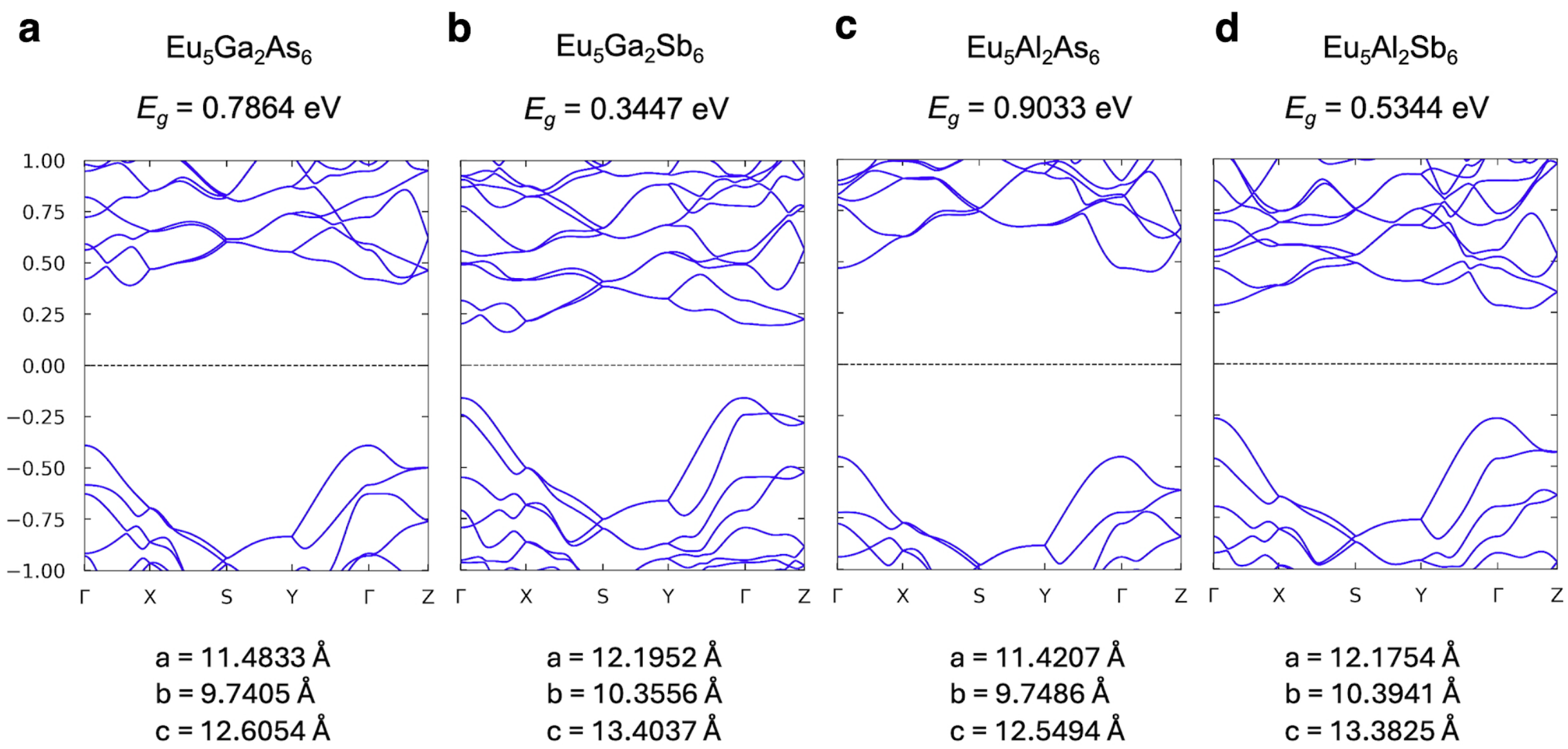

**Figure S11. Ab initio calculations of $Eu_5A_2Pn_6$ (A = Al, Ga; Pn = As, Sb) in *Pnma* space group phase. (a–d)** Electronic band structures of $Eu_5Ga_2As_6$, $Eu_5Ga_2Sb_2$, $Eu_5Al_2As_6$, and $Eu_5Al_2Sb_6$ in the paramagnetic order, respectively, showing calculated band gaps and lattice constants for each orthorhombic structure.

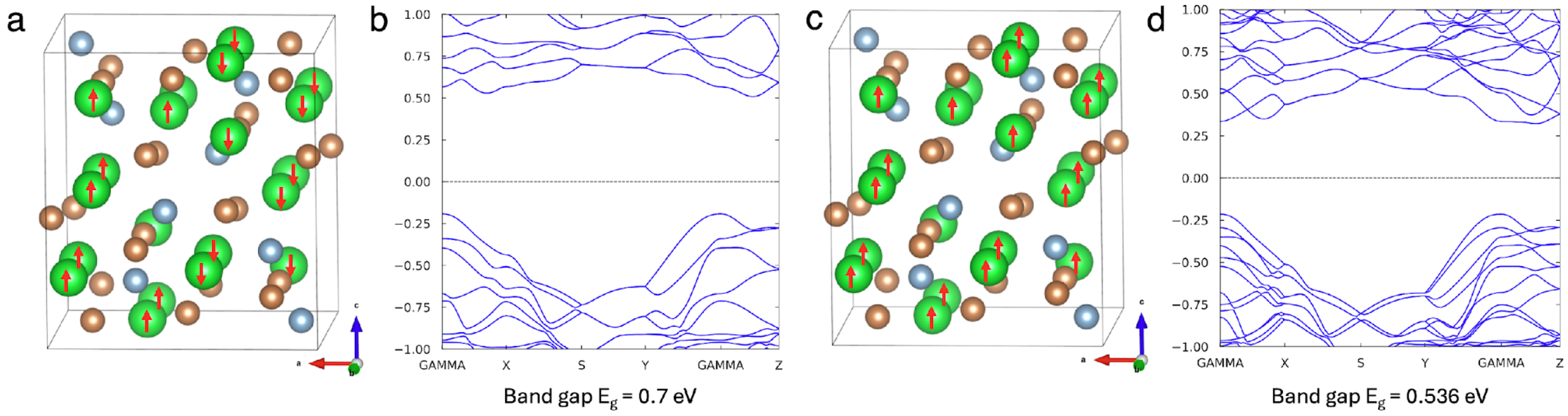

**Figure S12. Ab initio calculations of $Eu_5Ga_2As_6$ in antiferromagnetic and ferromagnetic orders. (a, c)** Crystal and magnetic structures of $Eu_5Ga_2As_6$ in antiferromagnetic (AFM) and ferromagnetic (FM) order, respectively. **(b, d)** Electronic band structures and band gaps of $Eu_5Ga_2As_6$ in AFM and FM orders, respectively. The AFM phase is favored ($E_{AFM} - E_{FM} = -2.2$ meV/Eu).

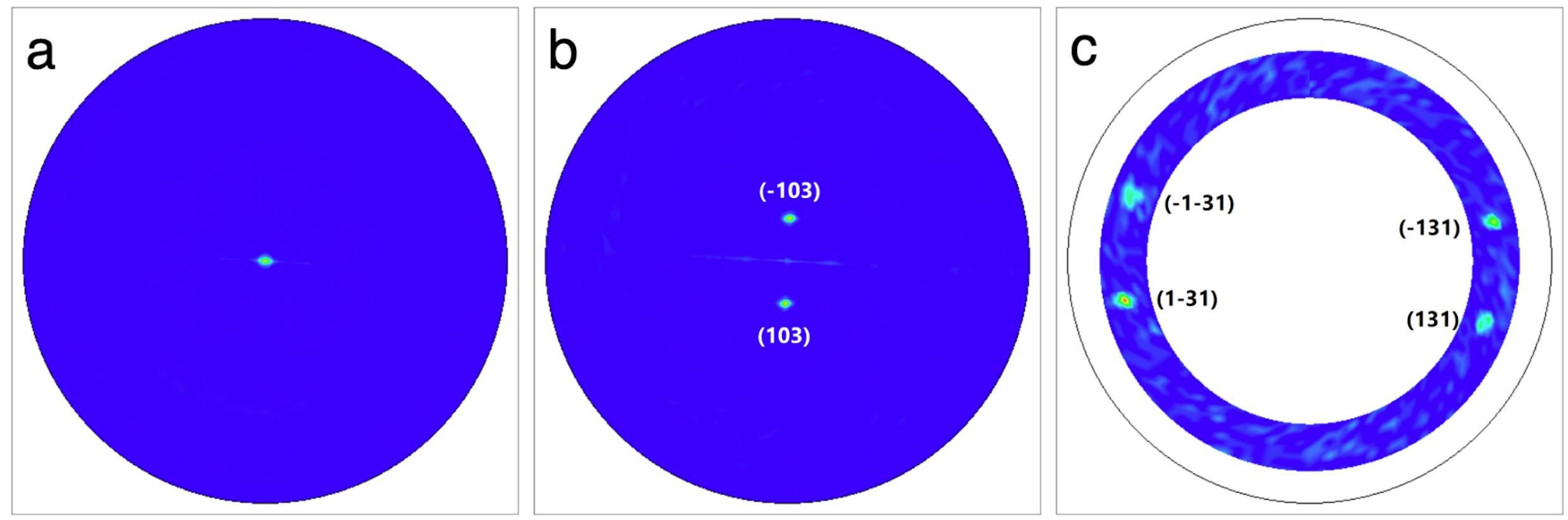

**Figure S13. XRD pole figure analyses of $Eu_5Ga_2As_6$ thin films on (001) GaAs substrates. (a)** Pole figure of the (001) plane, showing a single central spot corresponding to the major orientation of the film, as observed in θ-2θ scan. **(b)** Pole figure of the {103} planes, displaying two bright spots corresponding to the (103) and $(\underline{1}03)$ reflections, which appear due to the ~20° inclination of these planes relative to (001) in the Pnma structure. **(c)** Pole figure of the {131} planes, which are tilted by ~76° with respect to (001), showing the expected four spots. All pole figures are presented on a linear scale for improved clarity, both with and without indices.

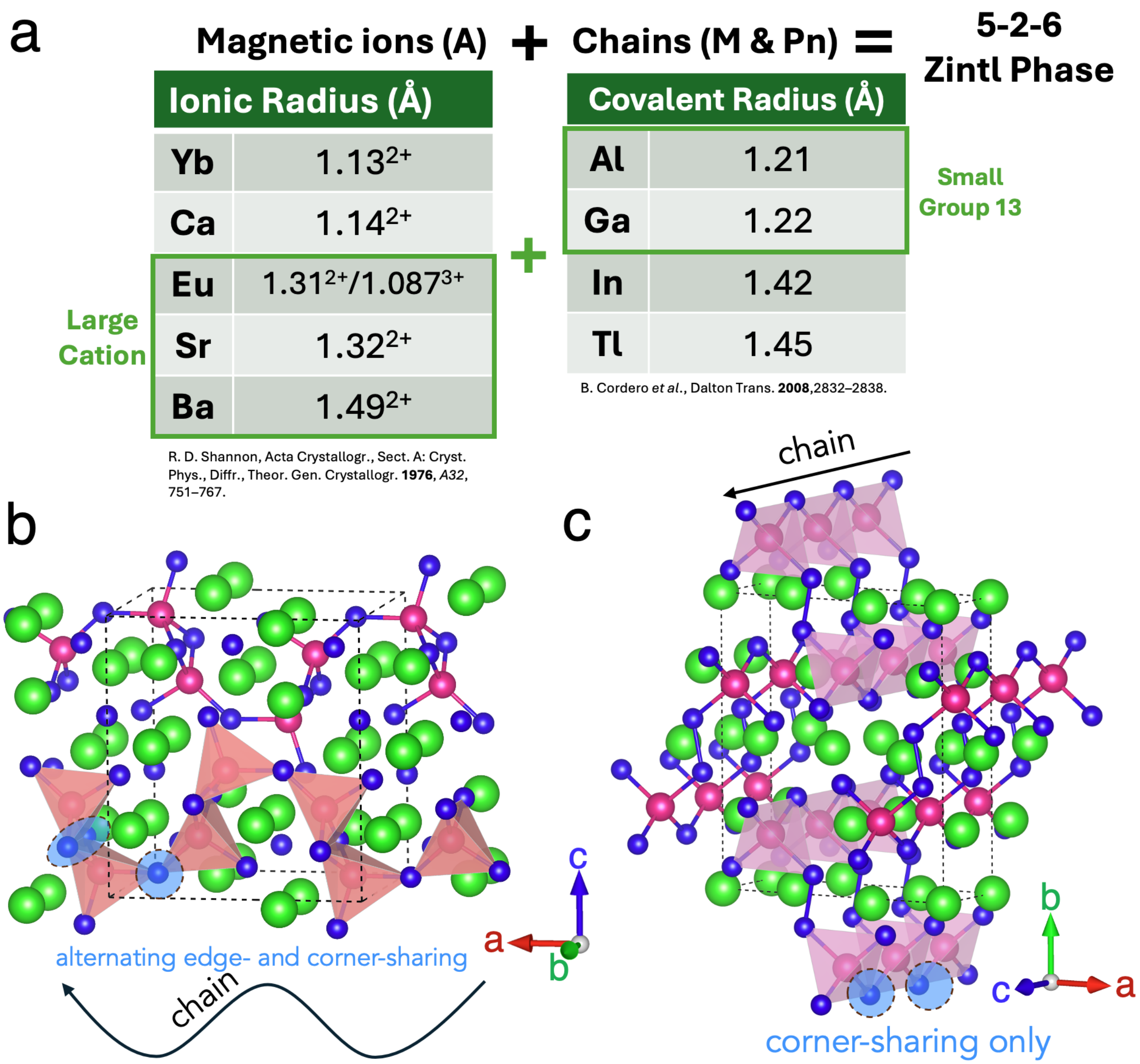

**Figure S14.** (a) Ionic radii of the A-site cations (A = Yb, Ca, Eu, Sr, Ba; Shannon ionic radii [Shannon1976]) and covalent radii of the Group 13 M-site elements (M = Al, Ga, In, Tl; Cordero covalent radii [Cordero2008]) used to evaluate the radius ratio $r_A/r_M$ in Table 2. The combination of a large A-site cation (Eu, Sr, and Ba) with a small Group 13 element (Al and Ga) yields the relatively low $r_A/r_M$ values associated with the Pnma structure. (b) Crystal structure of the $Sr_5Al_2Sb_6$-type *Pnma* phase viewed along the b axis. The polyanionic (M–Pn) chains are twisted, with alternating edge- and corner-sharing $MPn_4$ tetrahedra (edge-shared Pn atoms highlighted in blue); representative tetrahedra are shaded. (c) Crystal structure of the $Ca_5Ga_2As_6$-type *Pbam* phase viewed along the c axis, in which the chains consist of corner-sharing tetrahedra only (corner-shared Pn atoms highlighted in blue). A (green), M (magenta), and Pn (blue) atoms are shown; dashed lines mark the unit cells. Note that the Pnma unit cell is approximately twice the volume of the *Pbam* unit cell [Liu2015].

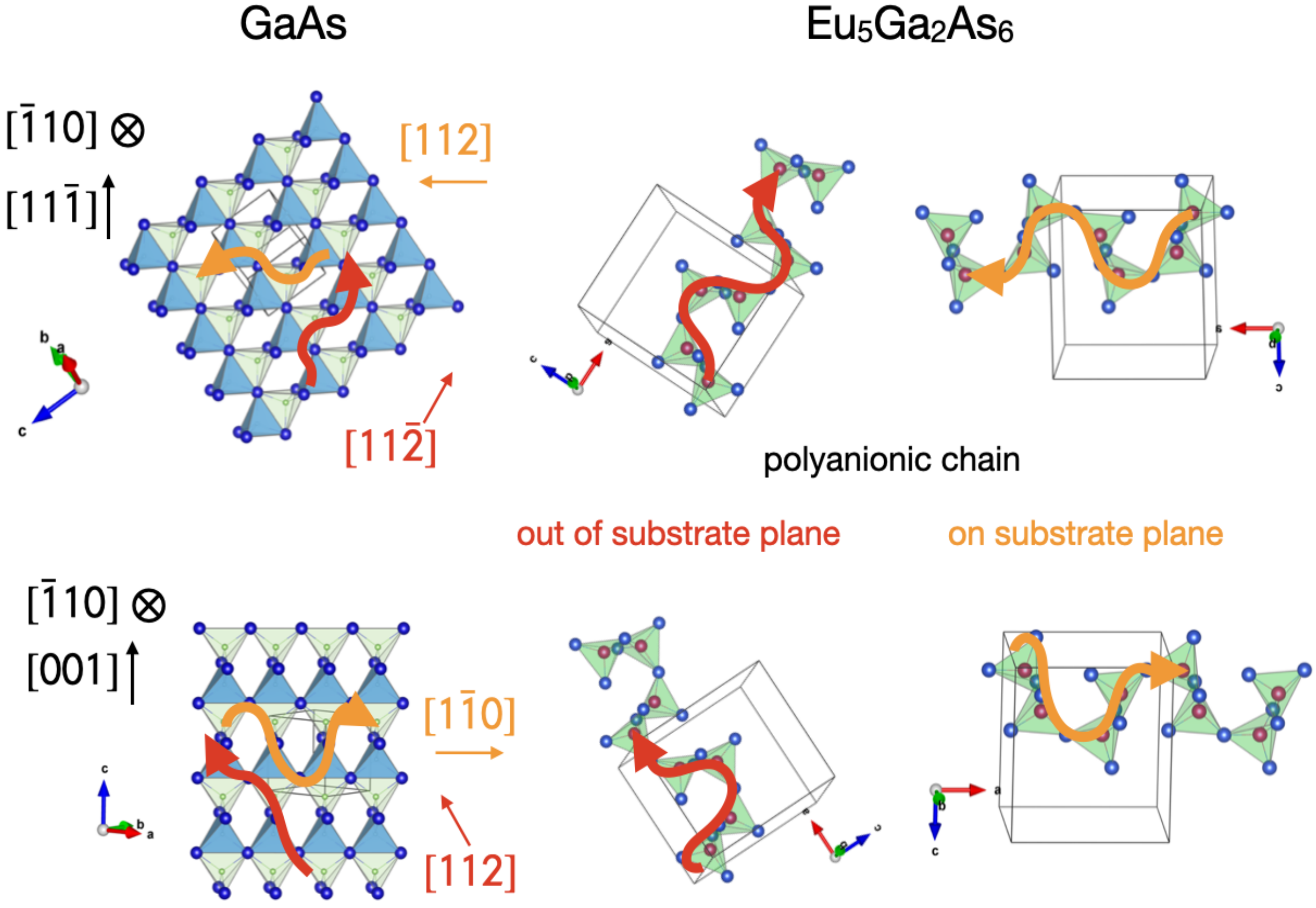

**Figure S15.** Alignment of the polyanionic chains in the Zintl phases with respect to the substrate. The polyanionic chains alternate through corner and edge-connected tetrahedra. The template for these chains is tetrahedral sites in the zinc-blend structure on the {110} planes connected along the <110> or <112> directions, resulting in polyanionic chains aligned inclined with respect to the trivial zinc-blend substrate surfaces.

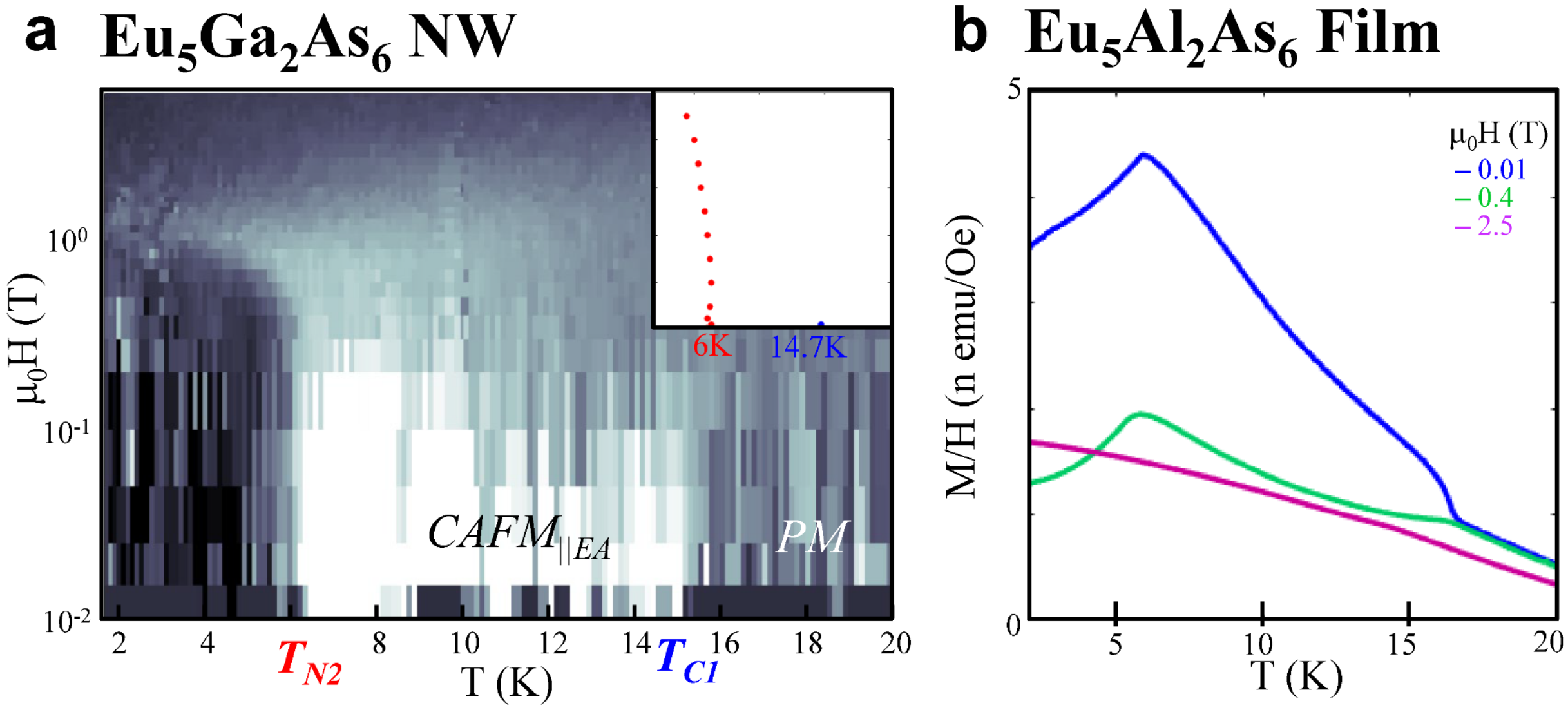

**Figure S16. Additional magnetization characterization data. a** Individual magnetization measurements of a $Eu_5Al_2As_6$ thin film along temperature sweeps at distinct applied magnetic fields. **b** Temperature-magnetic field phase diagram of $Eu_5Ga_2As_6$ NWs dispersed over a $Si/SiO_2$ substrate shown through dM/dT in false color. The inset shows the extracted transition lines.